\documentclass[twoside,english,eqsecnum,nofootinbib,superscriptaddress]{revtex4-2}
\usepackage[T1]{fontenc}
\usepackage{color}
\usepackage{babel}
\usepackage{array}
\usepackage{float}
\usepackage{mathtools}
\usepackage{bm}
\usepackage{multirow}
\usepackage{amsmath}
\usepackage{amssymb}
\usepackage{graphicx}
\usepackage{geometry}
\usepackage[pdfusetitle,
 bookmarks=true,bookmarksnumbered=false,bookmarksopen=false,
 breaklinks=false,pdfborder={0 0 1},backref=false,colorlinks=true]
 {hyperref}

\makeatletter

\providecommand{\tabularnewline}{\\}

\makeatother

\begin{document}

\title{Interacting hypersurfaces and multiple scalar-tensor theories}

\author{Yang Yu}
\affiliation{School of Physics and Astronomy, Sun Yat-sen University, Zhuhai 519082,
China}

\author{Zheng Chen}
\affiliation{School of Physics and Astronomy, Sun Yat-sen University, Zhuhai 519082,
China}

\author{Yu-Min Hu}
\affiliation{Department of Astronomy, School of Physical Sciences, University of Science and Technology of China, 96 Jinzhai Road, Hefei, Anhui 230026, China}
\affiliation{CAS Key Laboratory for Research in Galaxies and Cosmology, School of Astronomy and Space Science,
	University of Science and Technology of China, 96 Jinzhai Road, Hefei, Anhui 230026, China}

\author{Xian Gao}
\email[Corresponding author: ]{gaoxian@mail.sysu.edu.cn}
\affiliation{School of Physics and Astronomy, Sun Yat-sen University, Zhuhai 519082,
China}

\begin{abstract}
	We propose a novel method to construct ghost-free multiple scalar-tensor theories. The key idea is to use the geometric quantities of hypersurfaces defined by the scalar fields, rather than the covariant derivatives of scalar fields or spacetime curvature, to build the theory. This approach has proven effective in developing ghost-free scalar-tensor theories in the single-field case. When multiple scalar fields are present, each field specifies a foliation of spacelike hypersurfaces, on which we can define the normal vector, induced metric, extrinsic and intrinsic curvatures, as well as extrinsic (Lie) and intrinsic (spatial) derivatives, respectively. By employing these hypersurface geometric quantities as foundational elements, we construct the Lagrangian for interacting hypersurfaces that describes a multiple scalar-tensor theory. Given that temporal (Lie) and spatial derivatives are separated, it becomes relatively easier to control the order of time derivatives, thus helping to avoid ghost-like or unwanted degrees of freedom. In this work, we use bi-scalar-field theory as an example, focusing on polynomial-type Lagrangians. We construct monomials of hypersurface geometric quantities up to $d=3$, where $d$ denotes the number of derivatives in each monomial. Additionally, we present the correspondence between expressions in terms of hypersurface quantities and those in covariant bi-scalar-tensor theory. Through a cosmological perturbation analysis of a simple model, we demonstrate that the theory propagates two tensor and two scalar degrees of freedom at the linear order in perturbations, thereby remaining free from any extra degrees of freedom.
\end{abstract}

\maketitle

\section{Introduction}

As one of the most popular extensions to Einstein's general relativity (GR), scalar-tensor theories of gravity have garnered considerable research interest. By introducing a scalar field that is covariantly and non-minimally coupled with curvature terms, an additional scalar degree of freedom (DOF) beyond the two tensor modes is introduced, leading to a total of $2+1$ DOFs in these theories. This additional degree of freedom can result in phenomena that differ significantly from GR predictions \citep{Koyama:2015vza,Ferreira:2019xrr,Arai:2022ilw}.

The origins of scalar-tensor theories can be traced back to the formulation of $k$-inflation \citep{Armendariz-Picon:1999hyi}, which extended potential-dominated inflation theories to include first derivatives of the scalar field. Subsequent developments led to theories incorporating non-linear, second-order derivatives of the scalar field in a Minkowski background, known as Galileon theory, which respects Galileon symmetry \citep{Nicolis:2008in}. The theory was then extended to curved spacetime by introducing non-minimally coupled curvature terms \citep{Deffayet:2009wt,Deffayet:2009mn}. This generalized Galileon theory \citep{Deffayet:2011gz} was shown to be equivalent to Horndeski theory \citep{Horndeski:1974wa}. Horndeski's framework, formulated in 1974, represents the most general scalar-tensor Lagrangian with at most second-order derivatives that produces healthy, second-order field equations. Later developments included the degenerate higher-order derivative theory \citep{Gleyzes:2014dya,Gleyzes:2014qga,Langlois:2015cwa,Motohashi:2016ftl}, which further mitigates the Ostrogradsky ghost issue \citep{Woodard:2015zca} through degeneracy conditions, broadening the scope of healthy single-scalar-tensor theories. This naturally raises the question of whether we can extend these concepts to bi- or even multi-scalar-tensor theories.

Inspired by the development of single-scalar-tensor theories, numerous studies have aimed to formulate multi-scalar-tensor theories, with researchers seeking an analogue to Horndeski theory for multiple scalar fields. Multi-scalar theories with second-order field equations in a Minkowski background and respecting Galileon shift symmetry have been investigated \citep{Deffayet:2010zh,Padilla:2010de,Padilla:2010tj,Hinterbichler:2010xn,Padilla:2010ir} under the umbrella of multi-Galileon theory. Upon covariantizing the derivatives and introducing Riemann curvature coupling terms to account for commutation of derivatives, the multi-Galileon theory was initially extended to curved spacetime \citep{Padilla:2012dx}. It was later observed that the multi-Dirac-Born-Infeld Galileon theory includes additional terms beyond those in multi-Galileon theory \citep{Kobayashi:2013ina}. The generalized multi-Galileon theory was subsequently proposed \citep{Akama:2017jsa}, though it has been argued that this theory may suffer from geodesic incompleteness or gradient instabilities.

The most general second-order field equations for bi-scalar-tensor (BST) theories were formulated in \citep{Ohashi:2015fma} following an approach similar to Horndeski's original construction of single-scalar-tensor theories. Notably, single-scalar-tensor theories minimally coupled to an additional scalar matter field can be viewed as a subclass of BST theories, provided that the coupling preserves the total DOF count in the absence of matter \citep{Deffayet:2020ypa,Takahashi:2022ctx}. For instance, the theoretical framework in \citep{Gergely:2014rna} presents two propagating scalar degrees of freedom: one enters the gravitational sector, serving as a candidate for dark energy, while the other represents dark matter in unitary gauge. 
Recently, Horndeski attempted to construct the most general bi-scalar-tensor theory with second-order field equations using the traditional approach based on the trace of general second-order field equations \citep{Horndeski:2024hee}\footnote{Nevertheless, the original proof in \citep{Horndeski:2024hee} appears to be incomplete. We would like to thank Tsutomu Kobayashi for bringing this to our attention.}.
This naturally invites further exploration of bi-scalar-tensor theories beyond Horndeski and healthy scalar-tensor theories with more scalar fields. In this work, we introduce an alternative method for systematically constructing multi-field theories.

Non-generally covariant theories tend to possess additional DOFs beyond the two tensorial DOFs of GR. When a timelike scalar field is present, time reparametrization symmetry is broken, and scalar-tensor theories can be expressed as spatially covariant theories after gauge-fixing with $\phi=\phi(t)$, or unitary gauge, which encodes the scalar DOF in spatial geometric quantities. Alternatively, one can apply a gauge-recovering procedure to spatially covariant gravity theories to restore general covariance. Viewing spatially covariant gravity as a mediator in this construction process provides a starting point for systematically deriving general scalar-tensor theories.

Spatially covariant gravity \citep{Gao:2014soa,Gao:2014fra} refers to modified gravity theories that respect spatial diffeomorphism. Examples of such theories include effective field theories for inflation \citep{Creminelli:2006xe,Cheung:2007st} and dark energy \citep{Gubitosi:2012hu}, as well as Ho\v{r}ava gravity \citep{Horava:2009uw,Blas:2009qj}. The separation of space and time within these theories allows for greater flexibility in constructing ghost-free theories with specific DOFs. For instance, using elements from spatially covariant gravity, one can construct minimally modified gravity with the same DOFs as GR, as discussed in \citep{Gao:2019twq,Hu:2021yaq,Yao:2020tur,Yao:2023qjd,Wang:2024hfd}. Other studies on SCG theories can be found in \citep{Fujita:2015ymn,Saitou:2016lvb,Gao:2019liu,Zhu:2022dfq,Zhu:2022uoq,Zhu:2023rrx,Yu:2024drx,Hu:2024hzo}. Ghost-free spatially covariant gravity with 3 DOFs \citep{Fujita:2015ymn,Gao:2018znj,Gao:2019lpz,Gao:2018izs} relates closely to single-scalar-tensor theories, serving as a ``generator'' for scalar-tensor theories with higher derivatives, regardless of unitary gauge \citep{Gao:2020qxy,Gao:2020yzr} or not \citep{Hu:2021bbo}. Given these successful constructions, we are inspired to pursue an analogous approach for multi-scalar-tensor theories with higher derivatives, prompting us to first consider multi-field spatially covariant gravity.

In this work, we propose a method for constructing ghost-free multi-scalar-tensor theories when a unitary gauge for two fields is accessible, utilizing a correspondence between spatially covariant and generally covariant scalar-tensor theories. Specifically, we generalize spatially covariant gravity by introducing two independent foliation structures, with spatial geometric quantities derived from different timelike normal vectors associated with the two scalar fields. We use these geometric quantities to formulate the Lagrangian for interacting hypersurfaces that describe a bi-scalar-tensor theory and outline possible monomial terms for specific dimensions. We then extend the covariant correspondence procedure developed for the single-field case to the two-scalar-field scenario, applying it to derive generally covariant bi-scalar-tensor theories. We provide a specific example for $d=2$, examining its degrees of freedom (DOFs) using cosmological perturbation theory.

This paper is organized as follows. In Sec. \ref{sec:geo}, we introduce the geometric framework and quantities used in our construction. In Sec. \ref{sec:action}, we outline the procedure for constructing the Lagrangian for interacting hypersurfaces that describe a bi-scalar-tensor theory. In Sec. \ref{sec:mono}, we present all possible monomials for the Lagrangian up to $d=3$, with $d$ representing the total number of derivatives in each monomial. In Sec. \ref{sec:corr}, we apply the covariant correspondence procedure to obtain ghost-free bi-scalar-tensor theories with timelike scalar fields. In Sec. \ref{sec:pert}, we illustrate a $d=2$ bi-scalar-tensor theory and perform a perturbation analysis to further investigate the physical DOFs. Finally, in Sec. \ref{sec:Conclusion}, we summarize our discussion.

\textit{Notations:} Throughout this paper, we adopt the unit $8\pi G=1$ and use the metric convention $\{-,+,+,+\}$. Indices $\{a,b,c,...\}$ represent 4-dimensional coordinates, while indices $\{i,j,k,...\}$ refer to spatial components. $R$ denotes spacetime curvature, and $^{3}\!R$ denotes spatial curvature.

\section{Geometry of two foliations of hypersurfaces \label{sec:geo}}

\begin{figure}[h]
\begin{centering}
\includegraphics{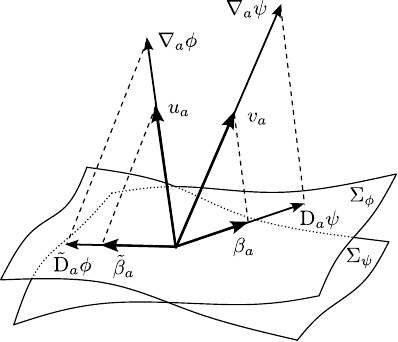}
\par\end{centering}
\caption{The two foliation structure specified by the two scalar fields $\phi$
and $\psi$ and their associated geometric quantities.}
\label{fig:twofoli}
\end{figure}

In this section, we introduce the geometry of two foliations of spacelike hypersurfaces. The basic concept of our construction is that two scalar fields exist in spacetime, denoted by $\phi$ and $\psi$, respectively. Each scalar field corresponds to a foliation of hypersurfaces on which the value of the scalar field is uniform. For our purposes, we assume that the gradients of the two scalar fields are both timelike. We denote the spacelike hypersurfaces corresponding to uniform $\phi$ (or uniform $\psi$) by $\Sigma_{\phi}$ (or $\Sigma_{\psi}$), respectively. The two foliations are accompanied by their own geometric quantities, as depicted in Fig. \ref{fig:twofoli}. 

We take the gradients of the scalar fields as normal vectors to the hypersurfaces, i.e., 
\begin{equation}
u_{a}\coloneqq -N\nabla_{a}\phi,\quad v_{a}\coloneqq -M\nabla_{a}\psi,
\end{equation}
where $N$ and $M$ are ``normalization factors'' such that $u_{a}$
and $v_{a}$ are normalized, i.e., $u^{a}u_{a}=v_{a}v^{a}=-1$. In
the case of a single scalar field, the normalization factor is equivalent
to the lapse function in the Arnowitt-Deser-Misner (ADM) coordinates.
We thus have
\begin{equation}
N=\frac{1}{\sqrt{2X}},\quad M=\frac{1}{\sqrt{2Y}},
\end{equation}
where $X$ and $Y$ are the canonical kinetic terms for each scalar
field defined by
\begin{equation}
X\coloneqq-\frac{1}{2}\nabla_{a}\phi\nabla^{a}\phi,\quad Y\coloneqq-\frac{1}{2}\nabla_{a}\psi\nabla^{a}\psi,\label{XY_def}
\end{equation}
respectively. For later convenience, we define the ``mixing'' kinetic
term
\begin{equation}
Z\coloneqq-\frac{1}{2}\nabla_{a}\phi\nabla^{a}\psi.\label{Z_def}
\end{equation}

On spatial hypersurfaces specified by the scalar fields, i.e., $\Sigma_{\phi}$
and $\Sigma_{\psi}$, the induced metrics (or projection tensors)
are defined as usual,
\begin{equation}
h_{ab}=g_{ab}+u_{a}u_{b},\quad l_{ab}=g_{ab}+v_{a}v_{b}.
\end{equation}

In the single scalar field case, the covariant derivative $\nabla_{a}$ is split into a timelike part in terms of the Lie derivative with respect to the normal vector and a spacelike part in terms of the covariant derivative adapted to the induced metric. Since we now have two foliations, we have two choices for splitting the covariant derivative into temporal and spatial parts. There are two temporal derivatives, chosen to be the Lie derivatives with respect to the normal vectors $\pounds_{\bm{u}}$ and $\pounds_{\bm{v}}$, respectively. Similarly, the two spatial derivatives are chosen as $\mathrm{D}_{a}$ and $\tilde{\mathrm{D}}_{a}$, which are the covariant derivatives projected by the induced metrics $h_{ab}$ and $\tilde{h}_{ab}$, respectively.

On each hypersurface, there are ``adapted'' intrinsic and extrinsic curvatures. The intrinsic Ricci tensors on $\Sigma_{\phi}$ and $\Sigma_{\psi}$ are given by $^{3}\!R_{ab}\equiv{}^{3}\!R_{ab}(h)$ and $^{3}\!\tilde{R}_{ab}\equiv{}^{3}\!\tilde{R}_{ab}(l)$, respectively. The extrinsic curvatures of $\Sigma_{\phi}$ and $\Sigma_{\psi}$ are defined by
\begin{equation}
K_{ab}=\frac{1}{2}\pounds_{\bm{u}}h_{ab},\quad L_{ab}=\frac{1}{2}\pounds_{\bm{v}}l_{ab},
\end{equation}
respectively.

Note generally $u_{a}$ and $v_{a}$ are not parallel to each other.
We may decompose $v_{a}$ with respect to $u_{a}$ as
\begin{equation}
v_{a}=-\alpha u_{a}+\beta_{a},\quad\text{with}\;u^{a}\beta_{a}\equiv0.
\end{equation}
Conversely, we can decompose $u_{a}$ with respect to $v_{a}$ as
\begin{equation}
u_{a}=-\tilde{\alpha}v_{a}+\tilde{\beta}_{a},\quad\text{with}\;v^{a}\tilde{\beta}_{a}=0.
\end{equation}
It is easy to check that
\begin{equation}
\alpha=\tilde{\alpha}=u_{a}v^{a},\label{alpha}
\end{equation}
and
\begin{equation}
\beta_{a}=-M\mathrm{D}_{a}\psi,\quad\tilde{\beta}_{a}=-N\tilde{\mathrm{D}}_{a}\phi.\label{beta}
\end{equation}
Note $\alpha$ is related to the mixing kinetic term $Z$ defined
in (\ref{Z_def}) by
\begin{equation}
\alpha=-2ZMN\equiv-\frac{Z}{\sqrt{XY}}.
\end{equation}
Due to the normalization of $u_{a}$ and $v_{a}$, $\alpha$ (keep
in mind that $\alpha\equiv\tilde{\alpha}$), $\beta_{a}$ and $\tilde{\beta}_{a}$
are not independent, which are related by
\begin{equation}
\alpha=-\sqrt{1+\beta^{a}\beta_{a}}=-\sqrt{1+\tilde{\beta}^{a}\tilde{\beta}_{a}}.
\end{equation}

The derivatives of scalar fields are naturally encoded in the
above hypersurface geometrical quantities. For example, the first
order Lie (temporal) derivatives of the scalar fields with respect
to their own normal vectors (i.e., the normal vector to the hypersurface
specified by the scalar field itself) naturally correspond to the
``normalization factors'',
\begin{equation}
\pounds_{\bm{u}}\phi=u^{a}\nabla_{a}\phi\equiv\frac{1}{N},\quad\pounds_{\bm{v}}\psi=v^{a}\nabla_{a}\psi\equiv\frac{1}{M}.
\end{equation}
While the spatial derivatives of the ``adapted'' scalar fields identically
vanish, e.g.,
\begin{equation}
\text{D}_{a}\phi=h_{a}^{a'}\nabla_{a'}\phi=0,\quad\tilde{\mathrm{D}}_{a}\psi=l_{a}^{a'}\nabla_{a'}\psi=0.\label{spaD}
\end{equation}
At this point we emphasize that identities in (\ref{spaD}) (and in
fact all expressions in the above) are generally covariant expressions
and thus hold in any coordinate systems\footnote{In the literature, the 3+1 decomposition is often presented by fixing
the time coordinate to be the scalar field. Here we emphasize that
this is not a priori necessary and it is more convenient to make the
decomposition ``covariantly''. More importantly, since there are
two scalar fields in our construction, there is no natural choice of
the time coordinate as one of the two scalar fields.}.

It is also possible to take ``mixing'' derivatives of the scalar
fields. For example, taking Lie derivative of $\psi$ with respect
to $u_{a}$ yields
\begin{equation}
\pounds_{\bm{u}}\psi=u^{a}\nabla_{a}\psi=-u^{a}\frac{v_{a}}{M}\equiv-\frac{\alpha}{M}. \label{LieD_u_psi}
\end{equation}
Similarly, we have
\begin{equation}
\pounds_{\bm{v}}\phi=-\frac{\alpha}{N}.
\end{equation}
It is interesting to note that the mixing kinetic term $Z\equiv-\frac{\alpha}{2MN}$
naturally arises from the mixing Lie derivatives. For the mixing spatial
derivatives, (\ref{beta}) yields
\begin{equation}
\mathrm{D}_{a}\psi=-\frac{1}{M}\beta_{a}=-\frac{1}{M}\left(v_{a}+\alpha u_{a}\right),
\end{equation}
and
\begin{equation}
\tilde{\mathrm{D}}_{a}\phi=-\frac{1}{N}\tilde{\beta}_{a}=-\frac{1}{N}\left(u_{a}+\alpha v_{a}\right). \label{tD_phi}
\end{equation}
As a result, all derivatives of $\phi$ and $\psi$ can be expressed
in terms of $N$, $M$, $u_{a}$ and $v_{a}$ and their derivatives.

The relevant geometric quantities are summarized in Table \ref{tab:blocks}. These quantities will serve as the basic building blocks in our construction of the two-field scalar-tensor theory.
\begin{table}[h]
\renewcommand\arraystretch{1.5}
\begin{centering}
\begin{tabular}{|c|c|c|}
\hline 
Geometric quantities & w.r.t. $\Sigma_{\phi}$ & w.r.t. $\Sigma_{\psi}$\tabularnewline
\hline 
Normal vectors & $u_{a}\coloneqq-N\nabla_{a}\phi$ & $v_{a}\coloneqq-M\nabla_{a}\psi$\tabularnewline
\hline 
Induced metrics & $h_{ab}\coloneqq g_{ab}+u_{a}u_{b}$ & $l_{ab}\coloneqq g_{ab}+v_{a}v_{b}$\tabularnewline
\hline 
Normalization factors & $N\coloneqq1/\sqrt{2X}$ & $M\coloneqq1/\sqrt{2Y}$\tabularnewline
\hline 
Accelerations & $a_{a}\coloneqq\text{D}_{a}\ln N$ & $b_{a}\coloneqq\tilde{\mathrm{D}}_{a}\ln M$\tabularnewline
\hline 
Extrinsic curvatures & $K_{ab}\coloneqq\frac{1}{2}\pounds_{\bm{u}}h_{ab}$ & $L_{ab}\coloneqq\frac{1}{2}\pounds_{\bm{v}}l_{ab}$\tabularnewline
\hline 
Ricci tensors & $^{3}\!R_{ab}\equiv{}^{3}\!R_{ab}(h)$ & $^{3}\!\tilde{R}_{ab}\equiv{}^{3}\!\tilde{R}_{ab}(l)$\tabularnewline
\hline 
\end{tabular}
\par\end{centering}
\caption{\label{tab:blocks}Basic geometric quantities of two foliations of
spacelike hypersurfaces.}
\end{table}

Up to now, it seems we have simply created two copies of geometric quantities for each foliation of hypersurfaces. Nevertheless, we would like to mention a subtle point in our construction. In the case of a single scalar field, there is a single normal vector, allowing all tensors to have a ``unique'' decomposition into their temporal and spatial parts. In our case, since there are two sets of hypersurfaces, we always have two choices for the decomposition of tensor fields. In particular, a tensor field that is ``tangent'' to $\Sigma_{\phi}$ (i.e., is spacelike and thus has no temporal part with respect to $\Sigma_{\phi}$) will contain a nonvanishing temporal part with respect to $\Sigma_{\psi}$ (i.e., is not a purely spatial tensor with respect to $\Sigma_{\psi}$), and vice versa.

For example, the induced metric $l_{ab}$ on $\Sigma_{\psi}$, which
is a spatial tensor with respect to $\Sigma_{\psi}$, can be further
decomposed with respect to $\Sigma_{\phi}$ as \citep{Hu:2021bbo}
\begin{equation}
l_{ab}=u_{a}u_{b}\beta_{c}\beta^{c}-2\alpha u_{(a}\beta_{b)}+h_{ab}+\beta_{a}\beta_{b},\label{lab_dec}
\end{equation}
in which $\alpha$ and $\beta_{a}$ are defined in (\ref{alpha})
and (\ref{beta}). Here $h_{ab}+\beta_{a}\beta_{b}$ is the spatial
(tangent) part of $l_{ab}$ with respect to $\Sigma_{\phi}$, while
$u_{a}u_{b}\beta_{c}\beta^{c}-2\alpha u_{(a}\beta_{b)}$ contains
temporal (normal) part with respect to $\Sigma_{\psi}$. Similarly,
for the acceleration $b_{a}$ on $\Sigma_{\psi}$, which is a vector
field tangent to $\Sigma_{\psi}$, we have \citep{Hu:2021bbo}
\begin{equation}
b_{a}=-u_{a}b_{\bm{u}}+b_{\hat{a}},\label{ba_dec}
\end{equation}
with
\begin{equation}
b_{\bm{u}}=-\beta^{c}\pounds_{\bm{u}}\beta_{c}+\alpha a^{c}\beta_{c}+\frac{1}{\alpha}\beta^{b}\beta^{c}\mathrm{D}_{b}\beta_{c},\label{acc_dec_1}
\end{equation}
and
\begin{equation}
b_{\hat{a}}=-\alpha\pounds_{\bm{u}}\beta_{a}+a_{a}\alpha^{2}+\beta^{b}\mathrm{D}_{b}\beta_{a}.\label{acc_dec_2}
\end{equation}
Here $b_{\hat{a}}$ is the spatial (tangent) part of $b_{a}$ with
respect to $\Sigma_{\phi}$, while $-u_{a}b_{\bm{u}}$ is the temporal
(normal) part. We refer to \citep{Hu:2021bbo} for more details on
the decomposition of geometric quantities on one hypersurface (such
as $\Sigma_{\psi}$) with respect to another (such as $\Sigma_{\phi}$)\footnote{In \citep{Hu:2021bbo}, the ``covariant 3+1 correspondences'' of
various geometric quantities are derived, which are the decomposition
of quantities in the unitary gauge (with normal vector $u_{a}$) with
respect to a general foliation (with normal vector $n_{a}$). For
our purpose to decompose quantities on $\Sigma_{\psi}$ with respect
to $\Sigma_{\phi}$, we just need to make the replacement $u_{a}\rightarrow v_{a}$
and $n_{a}\rightarrow u_{a}$ (and vice verse).}.

\section{Action for interacting hypersurfaces \label{sec:action}}

In the above, we have discussed the basic geometric quantities of the two foliations of spacelike hypersurfaces. We are now ready to construct the action based on these geometric quantities. For the foliation of $\Sigma_{\phi}$, the basic building blocks are the scalar field $\phi$, the normalization factor $N$, the normal vector $u_{a}$, the induced metric $h_{ab}$, and the intrinsic curvature of the hypersurface $^{3}\!R_{ab}$. Similarly, the building blocks for the foliation of $\Sigma_{\psi}$ are $\psi$, $M$, $v_{a}$, $l_{ab}$, and $^{3}\!\tilde{R}_{ab}$. All derivatives must be expressed in terms of the temporal and spatial derivatives $\pounds_{\bm{u}}$, $\pounds_{\bm{v}}$, $\mathrm{D}_{a}$, and $\tilde{\mathrm{D}}_{a}$, respectively.

The most general action built from these geometric quantities can be written as
\begin{equation}
S=\int\text{d}^{4}x\sqrt{g}\mathcal{L}\left(\phi,N,u_{a},h_{ab},{}^{3}\!R_{ab},\text{D}_{a},\pounds_{\bm{u}};\psi,M,v_{a},l_{ab},{}^{3}\!\tilde{R}_{ab},\tilde{\mathrm{D}}_{a},\pounds_{\bm{v}}\right),\label{S_gen}
\end{equation}
which thus describes two foliations of hypersurfaces. If the two foliations of hypersurfaces are decoupled from each other, i.e., if there is no interaction between them, then (\ref{S_gen}) reduces to
\begin{equation}
S=S_{\phi}+S_{\psi},
\end{equation}
where
\begin{equation}
S_{\phi}=\int\text{d}^{4}x\sqrt{g}\mathcal{L}\left(\phi,N,h_{ab},{}^{3}\!R_{ab},\text{D}_{a},\pounds_{\bm{u}}\right),
\end{equation}
and $S_{\psi}$ is similar to $S_{\phi}$ with $\phi$ replaced by
$\psi$. In this case the theory merely describes two decoupled scalar
fields. For our purpose, we are interested in building a theory in
which the two foliations of hypersurfaces interact with each other.
In principle, such interactions can be realized in two approaches,
one is through contraction between two sets of geometric quantities,
such as $u_{a}v^{a}$, $K_{ab}L^{ab}$ etc., the other is through
mixing derivatives such as $\pounds_{\bm{u}}l_{ab}$, $\mathrm{D}_{a}v_{b}$,
etc.

We have a couple of comments on the form of (\ref{S_gen}). First, since all derivatives of $\phi$ and $\psi$ can be expressed in terms of $N$, $M$, $u_{a}$, and $v_{a}$ and their derivatives, the explicit inclusion of the derivatives of the scalar fields $\phi$ and $\psi$ is unnecessary. Therefore, in (\ref{S_gen}), the Lie and spatial derivatives do not explicitly act on the scalar fields.

Second, in the case of a single scalar field, the explicit inclusion of the normal vector is unnecessary, as its contraction is either identically vanishing or a numerical constant. In our case, since $u_{a}$ and $v_{a}$ are not generally parallel, the contraction of $u_{a}$ with geometric quantities adapted to $\Sigma_{\psi}$ (and vice versa) will yield nontrivial contributions. Thus, we must keep $u_{a}$ and $v_{a}$ explicitly in the action (\ref{S_gen}).

Third, we assume that there should be no ``explicit'' mixing of derivatives. In other words, $\pounds_{\bm{u}}$ and $\mathrm{D}_{a}$ should only act on geometric quantities of $\Sigma_{\phi}$, while $\pounds_{\bm{v}}$ and $\tilde{\mathrm{D}}_{a}$ should only act on geometric quantities of $\Sigma_{\psi}$. Consequently, the interaction between the two sets of hypersurfaces can only occur through direct contractions of geometric quantities from the two foliations. This assumption is also consistent with the 3+1 decomposition of a two-scalar-field theory. We have at least two methods to perform the decomposition. One is to decompose all quantities with respect to a ``common'' foliation of hypersurfaces with a normal vector $n_{a}$, which is unrelated to $u_{a}$ or $v_{a}$. The other method is to decompose quantities associated with each scalar field according to the foliation of hypersurfaces defined by itself. The latter naturally corresponds to the action (\ref{S_gen}) but without any mixing derivatives\footnote{In fact, mixing derivative terms can always be recast as linear combinations of terms without mixing derivatives (see eqs. (\ref{LieD_u_psi})-(\ref{tD_phi}) as examples). This can also be understood since mixing derivative terms can always be rewritten in terms of expressions from a scalar-tensor theory, which can then be re-decomposed into terms without mixing derivatives.}.

We emphasize that the absence of mixing derivatives appearing explicitly in the Lagrangian does not mean that the two sets of hypersurfaces, or more precisely, the two scalar fields, are not coupled through derivatives. The mixing kinetic term $Z$ arises implicitly from couplings such as $u_{a}v^{a}$, $h_{ab}l^{ab}$, etc. The two scalar fields can also be coupled through higher derivatives. In fact, the mixing derivatives have already been encoded in the aforementioned geometric quantities and their ``adapted'' temporal/spatial derivatives. For example, the mixing kinetic term $Z$ naturally arises from
\begin{equation}
h^{ab}l_{ab}=2+\alpha^{2}=2+\frac{Z^{2}}{XY}.
\end{equation}
The key point is that these mixing derivatives should not appear ``explicitly'' in the Lagrangian but only implicitly in terms of contractions of geometric quantities from the two foliations.

Non-degenerate higher temporal derivatives will introduce extra degrees of freedom, which are typically ghostlike. In the case of a single scalar field, a class of theories with two tensor and one scalar degrees of freedom has been constructed \citep{Gao:2014soa}, which has been shown to be ghost-free through Hamiltonian analysis \citep{Gao:2014fra}. The idea is to introduce only first-order temporal (Lie) derivatives, which act only on the induced metric. That is, the temporal derivative enters the Lagrangian solely through the extrinsic curvature\footnote{It is also possible to introduce the time derivative of the lapse function, but with additional conditions \citep{Gao:2018znj}.}. In the case of two scalar fields, one might expect the theory to propagate two tensor and two scalar degrees of freedom. To achieve this, the idea is to construct a similar framework by requiring that the Lie derivatives enter the Lagrangian only via the two extrinsic curvatures $K_{ab}$ and $L_{ab}$. The resulting action thus takes the form
\begin{align}
S & =\int\text{d}^{4}x\sqrt{g}\mathcal{L}\left(\phi,N,u_{a},h_{ab},K_{ab},{}^{3}\!R_{ab},\mathrm{D}_{a};\psi,M,v_{a},l_{ab},L_{ab},{}^{3}\!\tilde{R}_{ab},\tilde{\mathrm{D}}_{a}\right).\label{S_fin}
\end{align}
Although a strict Hamiltonian analysis is still needed, it is reasonable to believe that (\ref{S_fin}) describes a ghost-free theory with two tensor and two scalar degrees of freedom.

In the case of a single scalar field, introducing only the extrinsic curvature (i.e., the Lie derivative of the induced metric) is merely a sufficient condition to guarantee that the theory is ghost-free. In principle, the Lie derivative of the lapse function can be introduced, but with additional conditions, ensuring that the theory propagates no more than three degrees of freedom \citep{Gao:2018znj, Lin:2020nro}. One might expect that (\ref{S_fin}) can be extended by including the first-order Lie derivatives of the normalization factors, i.e., $\pounds_{\bm{u}}N$ and $\pounds_{\bm{v}}M$, while maintaining the number of degrees of freedom unchanged. Again, some additional conditions must be imposed. Nevertheless, (\ref{S_fin}) provides a reasonable starting point for building the theory.

There is one subtlety in our construction. We now have two sets of geometric quantities, which are adapted to two foliations of hypersurfaces, respectively. According to the above discussion, since the two sets of geometric quantities are defined (as spatial or tangent tensors) with respect to different foliations, the spatial tensors on one foliation are no longer purely spatial tensors with respect to the other foliation. This can be seen explicitly from (\ref{lab_dec}) and (\ref{ba_dec}). More seriously, apparent higher temporal (Lie) derivatives (such as $\pounds_{\bm{u}}\beta_{c}$ in (\ref{acc_dec_1}) and (\ref{acc_dec_2})) may appear, potentially indicating the presence of extra degrees of freedom. Precisely, for an observer with 4-velocity $u^{a}$, both the temporal and spatial parts (with respect to $\Sigma_{\phi}$) of $b_{a}$, which represents the acceleration of another observer with velocity $v^{a}$, will contain seemingly dangerous parts, i.e., the second-order temporal derivative of $\psi$ evaluated from $\pounds_{\bm{u}}\beta_{a}$. This scenario is analogous to the case studied in single field theory, where the so-called ``instantaneous mode'' arises when deviating from the unitary gauge \citep{DeFelice:2018ewo, DeFelice:2021hps}. However, lessons from the construction of spatially covariant gravity and the corresponding scalar-tensor theory indicate that a spatial tensor (with respect to some spacelike hypersurfaces) containing no higher temporal derivatives in its ``own'' unitary gauge is always safe. In other words, as long as we can find a coordinate system in which an operator shows no higher temporal derivatives, such an operator can be safely included in the Lagrangian. This is the case for the geometric quantities associated with each foliation of hypersurfaces.

The above argument can be applied to multiple foliations of spacelike hypersurfaces. The point is that it is not possible to choose a unitary gauge for all scalar fields simultaneously. For $n$ scalar fields, we can choose a unitary gauge for at most one of them. For the remaining $n-1$ scalar fields, the corresponding geometric quantities in such a unitary gauge will inevitably contain higher temporal derivatives and introduce multiple (at least $n-1$) instantaneous modes. Nevertheless, as long as these quantities contain no higher temporal derivatives in their own unitary gauges, it is safe to include them.

\section{Monomials built of geometric quantities of hypersurfaces \label{sec:mono}}

To construct more explicit Lagrangians, we consider Lagrangians that are linear combinations of scalar contractions of the building blocks discussed in the preceding sections. For convenience, we refer to these scalar contractions as ``hypersurface monomials''\footnote{In \citep{Gao:2020yzr,Gao:2020qxy}, such monomials are referred to as ``spatially covariant gravity (SCG) monomials.''}.

Since the number of building blocks and monomials can be extensive, it is useful to classify them. We follow the method developed in \citep{Gao:2020yzr,Gao:2020qxy} (see also \citep{Gao:2020juc}). The basic idea is to establish a correspondence between a monomial built from the geometric quantities of hypersurfaces, i.e., a hypersurface monomial, and an expression in scalar-tensor theory (usually a scalar-tensor polynomial). We label each hypersurface monomial by a set of integers $\left(c_{0}, c_{1}, c_{2}, \cdots; d_{1}, d_{2}, d_{3}, \cdots\right)$, where $c_{n}$ represents the number of $n$-th order derivatives of the scalar field and $d_{n}$ represents the number of $n$-th order derivatives of the curvature tensor in the corresponding scalar-tensor expression. The total number of derivatives of the hypersurface monomial is thus given by
\begin{equation}
d=\sum_{n=0}\left[(n+2)c_{n}+(n+1)d_{n+2}\right].
\end{equation}
Since $d_{1}$ does not contribute to $d$, we can suppress it and denote the set of integers as $(c_{0}, c_{1}, \cdots; d_{2}, d_{3}, \cdots)$. In Table \ref{tab:bldblc}, we list the building blocks up to $d=2$, where only three integers $\left(c_{0}; d_{2}, d_{3}\right)$ are necessary. At this point, note that we have omitted the scalar fields $\phi$ and $\psi$ themselves in Table \ref{tab:bldblc}, since all derivatives of the scalar fields can be encoded in other hypersurface geometric quantities. In other words, in our construction, the scalar fields appear explicitly only without derivatives.

\begin{table}[H]
\renewcommand\arraystretch{1.5}
\begin{centering}
\begin{tabular}{|c|c|c|}
\hline 
$d$ & Building blocks & $\left(c_{0};d_{2},d_{3}\right)$\tabularnewline
\hline 
\hline 
0 & $N,\;u_{a},\;h_{ab},\;M,\;v_{a},\;l_{ab}$ & $\left(0;0,0\right)$\tabularnewline
\hline 
1 & $K_{ab},\;a_{a},\;L_{ab},\;b_{a}$ & $\left(0;1,0\right)$\tabularnewline
\hline 
\multirow{2}{*}{2} & $^{3}\!R_{ab},\;{}^{3}\!\tilde{R}_{ab}$ & $\left(1;0,0\right),\left(0;2,0\right)$\tabularnewline
\cline{2-3}
 & $\mathrm{D}_{c}K_{ab},\;\mathrm{D}_{a}a_{b},\;\tilde{\mathrm{D}}_{c}L_{ab},\;\tilde{\mathrm{D}}_{a}b_{b}$ & $\left(0;0,1\right),\left(0;2,0\right)$\tabularnewline
\hline 
\end{tabular}
\par\end{centering}
\caption{Classification of building blocks up to $d=2$.}
\label{tab:bldblc}
\end{table}

\subsection{$d=0$}

We are now ready to construct the hypersurface momomials. In the case
of a single scalar field, the only nontrivial monomial of $d=0$ is
the lapse function $N$, since the contractions of $u_{a}$ and $h_{ab}$
are numerical constants. In our case, there will be nontrivial contractions
among $u_{a}$, $v_{a}$, $h_{ab}$ and $l_{ab}$. Following \citep{Hu:2024hzo},
we denote $\left[\cdots\right]$ as ``unfactorizable'' contractions
among tensor fields, where the indices are contracted using the ``spacetime''
metric. For the sake of notational briefness, we use $u$, $v$, $h$,
$l$, to denote $u_{a}$, $v_{a}$, $h_{ab}$ and $l_{ab}$, respectively.
For example,
\begin{equation}
\left[uv\right]\coloneqq u^{a}v_{a}=\alpha,\label{ctr_uv}
\end{equation}
and
\begin{eqnarray}
\left[hv\right]_{a} & \coloneqq & h_{a}^{b}v_{b}=v_{a}+\alpha u_{a},\label{ctr_hv}\\
\left[lu\right]_{a} & \coloneqq & l_{a}^{b}u_{b}=u_{a}+\alpha v_{a}.\label{ctr_lu}
\end{eqnarray}
(\ref{ctr_hv}) and (\ref{ctr_lu}) implies that the contractions
of $u_{a}$ and $v_{a}$ with the induced metrics are always linear
combinations of $u_{a}$ and $v_{a}$, with $\alpha$ as the coefficient.

It is also interesting to note that there are nontrivial contractions
between $h_{ab}$ and $l_{ab}$. We have
\begin{eqnarray}
\left[hl\right]_{\phantom{a}b}^{a} & \coloneqq & h_{c}^{a}l_{b}^{c}=g_{b}^{a}+v^{a}v_{b}+u^{a}u_{b}+\alpha u^{a}v_{b},\\
\left[lh\right]_{\phantom{a}b}^{a} & \coloneqq & l_{c}^{a}h_{b}^{c}=g_{b}^{a}+u^{a}u_{b}+v^{a}v_{b}+\alpha v^{a}u_{b}.
\end{eqnarray}
Also
\begin{eqnarray}
\left[hlh\right]_{\phantom{a}b}^{a} & \coloneqq & h_{c}^{a}l_{d}^{c}h_{b}^{d}=g_{b}^{a}+\left(1+\alpha^{2}\right)u^{a}u_{b}+v^{a}v_{b}+\alpha u^{a}v_{b}+\alpha v^{a}u_{b},\\
\left[lhl\right]_{\phantom{a}b}^{a} & \coloneqq & l_{c}^{a}h_{d}^{c}l_{b}^{d}=g_{b}^{a}+\left(1+\alpha^{2}\right)v^{a}v_{b}+u^{a}u_{b}+\alpha v^{a}u_{b}+\alpha u^{a}v_{b}.
\end{eqnarray}
Fortunately, higher order contractions are not independent, as one
can check that
\begin{equation}
\left[hlhl\right]_{\phantom{a}b}^{a}=\left[hlh\right]_{\phantom{a}b}^{a}+\alpha^{2}\left(\left[hl\right]_{\phantom{a}b}^{a}-h_{b}^{a}\right),\label{hlhlred}
\end{equation}
and thus
\begin{equation}
\left[hlhlh\right]_{\phantom{a}b}^{a}=\left(1+\alpha^{2}\right)\left[hlh\right]_{\phantom{a}b}^{a}-\alpha^{2}h_{b}^{a},
\end{equation}
and so on.

As a result, monomials of $d=0$, which are scalar contractions built
of $u_{a}$, $v_{a}$, $h_{ab}$ and $l_{ab}$, are merely functions
of $\alpha$. For example, $\left[uv\right]=\alpha$ as given in (\ref{ctr_uv}),
\begin{equation}
\left[hvv\right]\equiv h_{ab}v^{a}v^{v}=-1+\alpha^{2},
\end{equation}
and
\begin{equation}
\left[luu\right]\equiv l_{ab}u^{a}u^{v}=-1+\alpha^{2}.
\end{equation}
As for the contractions between $h_{ab}$ and $l_{ab}$, one finds
\begin{equation}
\left[hl\right]\equiv h_{ab}l^{ab}=2+\alpha^{2}=\left[hlh\right]\equiv h_{b}^{a}l_{c}^{b}h_{a}^{c},
\end{equation}
and
\begin{equation}
\left[hlhl\right]\equiv h_{b}^{a}l_{c}^{b}h_{d}^{c}l_{a}^{d}=2+\alpha^{4},
\end{equation}
and so on.

To conclude, the monomials of $d=0$ are general functions of $N$,
$M$ and $\alpha$ (as well as the scalar fields $\phi$ and $\psi$).

\subsection{$d=1$}

As for monomials of $d=1$, the contractions can be made between the
set of building blocks of $d=1$, i.e., $\left\{ K_{ab},a_{a},L_{ab},b_{a}\right\} $,
and the set of building blocks of $d=0$, i.e., $\left\{ u_{a},h_{ab},v_{a},l_{ab}\right\} $,
with indices contracted by the spacetime metric. Here we do not need
to consider $N$ and $M$ since they have no index to be contracted.
We find 2 independent monomials involving $K_{ab}$\footnote{Note $\left[Kvv\right]=K_{ab}v^{a}v^{b}=K_{ab}\left(l^{ab}-g^{ab}\right)=\left[Kl\right]-\left[K\right]$,
$\left[Klh\right]=K_{b}^{a}l_{c}^{b}h_{a}^{c}=K_{b}^{c}l_{c}^{b}=\left[Kl\right]$,
and $\left[Klhl\right]=\left(1+\alpha^{2}\right)\left[Kl\right]-\alpha^{2}\left[K\right]$,
etc. According to (\ref{hlhlred}), contractions with higher order
of $h_{ab}$ and $l_{ab}$ are not independent.},
\begin{equation}
\left[K\right]\equiv\left[Kh\right]=K_{ab}h^{ab},\quad\left[Kl\right]=K_{ab}l^{ab},
\end{equation}
and 1 monomial involving $a_{a}$,
\begin{equation}
\left[av\right]=a_{a}v^{a}.
\end{equation}
Here and in the following, in $\left[\cdots\right]$ we use $K$ to
denote $K_{ab}$, $a$ to denote $a_{i}$, and $^{3}\!R$ to denote
$^{3}\!R_{ab}$, etc. Similarly, there are 2 independent monomials
involving $L_{ab}$,
\begin{equation}
\left[L\right]\equiv\left[Ll\right]=L_{ab}l^{ab},\quad\left[Lh\right]=L_{ab}h^{ab},
\end{equation}
and 1 monomial involving $b_{a}$,
\begin{equation}
\left[bu\right]=b_{a}u^{a}.
\end{equation}

At this point, note that since the induced metrics $h_{ab}$ and $l_{ab}$
are combinations of the spacetime metric $g_{ab}$ and bi-vectors
$u_{a}u_{b}$ and $v_{a}v_{b}$, the contractions with $h_{ab}$ and
$l_{ab}$ may be equivalently and more conveniently written in terms
of contractions with $u_{a}$ and $v_{a}$. For example,
\begin{align}
\left[Kl\right] & =K_{ab}\left(g^{ab}+v^{a}v^{b}\right)=\left[K\right]+\left[Kvv\right],\\
\left[Lh\right] & =L_{ab}\left(g^{ab}+u^{a}u^{b}\right)=\left[L\right]+\left[Luu\right].
\end{align}

The independent monomials of $d=1$ are summarized in Table \ref{tab:Mono1}\footnote{``B.b.'' is the shorthand for ``building blocks''.}.
Note that in the case of a single scalar field, it is not possible to construct
a scalar of the acceleration. However, in the case of two scalar fields,
we have nontrivial monomials $\left[av\right]$ and $\left[bu\right]$.
\begin{table}[H]
\renewcommand\arraystretch{1.5}
\begin{centering}
\begin{tabular}{|c|c|}
\hline 
B.b. & Monomials\tabularnewline
\hline 
$K$ & $\left[K\right],\left[Kvv\right]$\tabularnewline
\hline 
$L$ & $\left[L\right],\left[Luu\right]$\tabularnewline
\hline 
$a$ & $\left[av\right]$\tabularnewline
\hline 
$b$ & $\left[bu\right]$\tabularnewline
\hline 
\end{tabular}
\par\end{centering}
\caption{Monomials in $d=1.$}
\label{tab:Mono1}
\end{table}

\subsection{$d=2$}

Similarly, monomials of $d=2$ are listed in Table. \ref{tab:mono2}.
\begin{table}[H]
\renewcommand\arraystretch{1.5}
\begin{centering}
\begin{tabular}{|c|c|c|}
\hline 
B.b. & Unfactorizable monomials & Factorizable monomials\tabularnewline
\hline 
$KK$ & $\left[KK\right]$, $\left[KvKv\right]$  & $[K]^{2},[Kvv]^{2},[K][Kvv]$\tabularnewline
\hline 
$LL$ & $\left[LL\right]$, $\left[LuLu\right]$ & $[L]^{2},[Luu]^{2},[L][Luu]$\tabularnewline
\hline 
$KL$ & $\left[KL\right]$, $\left[KvLu\right]$ & $[K][L],[Kvv][L],[K][Luu],[Kvv][Luu]$\tabularnewline
\hline 
$Ka$ & $\left[Kva\right]$ & $[K][av],[Kvv][av]$\tabularnewline
\hline 
$Kb$ & $\left[Kvb\right]$ & $[K][bu],[Kvv][bu]$\tabularnewline
\hline 
$La$ & $\left[Lua\right]$ & $[L][av],[Luu][av]$\tabularnewline
\hline 
$Lb$ & $\left[Lub\right]$ & $[L][bu],[Luu][bu]$\tabularnewline
\hline 
$aa$ & $\left[aa\right]$ & $[av]^{2}$\tabularnewline
\hline 
$ab$ & $\left[ab\right]$ & $[av][bu]$\tabularnewline
\hline 
$bb$ & $\left[bb\right]$ & $[bu]^{2}$\tabularnewline
\hline 
$^{3}\!R$ & $\left[^{3}\!R\right]$, $\left[^{3}\!Rvv\right]$ & -\tabularnewline
\hline 
$^{3}\!\tilde{R}$ & $\left[^{3}\!\tilde{R}\right]$, $\left[^{3}\!\tilde{R}uu\right]$ & -\tabularnewline
\hline 
$\mathrm{D}K$ & $\left[v\mathrm{D}K\right]$, $\left[\mathrm{D}Kv\right]$, $\left[v\mathrm{D}Kvv\right]$ & -\tabularnewline
\hline 
$\tilde{\mathrm{D}}L$ & $\left[u\tilde{\mathrm{D}}L\right]$, $\left[\tilde{\mathrm{D}}Lu\right]$,
$\left[u\tilde{\mathrm{D}}Luu\right]$ & -\tabularnewline
\hline 
$\mathrm{D}a$ & $\left[\mathrm{D}a\right]$, $\left[v\mathrm{D}av\right]$ & -\tabularnewline
\hline 
$\tilde{\mathrm{D}}b$ & $\left[\tilde{\mathrm{D}}b\right]$, $\left[u\tilde{\mathrm{D}}bu\right]$ & -\tabularnewline
\hline 
\end{tabular}
\par\end{centering}
\caption{Monomials of $d=2$. A \textquotedblleft -\textquotedblright{} indicates
no monomial in such a category.}

\label{tab:mono2}
\end{table}

In the above, we define
\begin{eqnarray}
\left[KvKv\right] & \coloneqq & K_{b}^{a}v_{a}K_{c}^{b}v^{c},\\
\left[KvLu\right] & \coloneqq & K_{b}^{a}v_{a}L_{c}^{b}u^{c},\\
\left[LuLu\right] & \coloneqq & L_{b}^{a}u_{a}L_{c}^{b}u^{c},
\end{eqnarray}
\begin{eqnarray}
\left[v\mathrm{D}K\right] & \coloneqq & v^{c}\mathrm{D}_{c}K_{ab}h^{ab},\\
\left[\mathrm{D}Kv\right] & \coloneqq & h^{ca}\mathrm{D}_{c}K_{ab}v^{b},\\
\left[v\mathrm{D}Kvv\right] & \coloneqq & v^{c}\mathrm{D}_{c}K_{ab}v^{a}v^{b},
\end{eqnarray}
and
\begin{equation}
\left[v\mathrm{D}av\right]  \equiv v^{a}\mathrm{D}_{a}a_{b}v^{b},
\end{equation}
and the same as $\left[u\tilde{\mathrm{D}}L\right]$, $\left[u\tilde{\mathrm{D}}bu\right]$
etc. At this point, it is interesting to note that in the case of
a single field, there is no nontrivial monomials in the form of $\mathrm{D}K$
and $\mathrm{D}a$, since it is impossible to construct a scalar by contracting
$\mathrm{D}_{c}K_{ab}$ and $\mathrm{D}_{a}a^{a}$ is a total derivative. 
At this point, note $[\mathrm{D}a]$ and $\left[\tilde{\mathrm{D}}b\right]$ are total derivatives by themselves, which thus can be neglected at $d=2$. We keep them in Table \ref{tab:mono2} since they will contribute monomials (factorizable ones) at $d=3$ or higher.

\subsection{$d=3$}

We can also construct monomials of $d=3$. Due to the large number of monomials, we first concentrate on the unfactorizable monomials and split them into three groups. The first group involves building blocks of $d\geq1$ that belong to $\Sigma_{\phi}$ only. Similarly, the second group involves building blocks of $d\geq1$ that belong to $\Sigma_{\psi}$ only. The third group involves building blocks from both $\Sigma_{\phi}$ and $\Sigma_{\psi}$. The three groups of monomials are listed in Tables \ref{tab:mono3_1}, \ref{tab:mono3_2}, and \ref{tab:mono3_3}, respectively. We emphasize that such a splitting is merely technical. For example, due to the contraction with $v_{a}$, even in the first group, there are couplings between geometric quantities from $\Sigma_{\phi}$ and $\Sigma_{\psi}$. For completeness, we present the factorizable monomials in Appendix \ref{app:mono3}.
\begin{table}[H]
\renewcommand\arraystretch{1.5}
\begin{centering}
\begin{tabular}{|c|c|}
\hline 
B.b. & Unfactorizable monomials\tabularnewline
\hline 
$KKK$ & $[KKK]$, $[vKKKv]$\tabularnewline
\hline 
$KKa$ & $[vKKa]$\tabularnewline
\hline 
$Kaa$ & $[Kaa]$\tabularnewline
\hline 
$K^{3}\!R$ & $[K^{3}\!R]$, $[vK^{3}\!Rv]$\tabularnewline
\hline 
$K\text{D}K$ & $[vK\text{D}K]_{1}$, $[vK\text{D}K]_{2}$, $[Kv\text{D}K]$, $[K\text{D}Kv]$,
$[vK\text{D}Kvv]$, $[Kv\text{D}Kvv]$\tabularnewline
\hline 
$K\text{D}a$ & $[K\text{D}a]$, $[vKv\text{D}a]$, $[vK\text{D}av]$\tabularnewline
\hline 
$aaa$ & -\tabularnewline
\hline 
$a^{3}\!R$ & $[a^{3}\!Rv]$\tabularnewline
\hline 
$a\text{D}a$ & $[av\text{D}a]$, $[a\text{D}av]$\tabularnewline
\hline 
$\text{D}\text{D}K$ & $[\text{D}\text{D}K]_{1}$, $[\text{D}\text{D}K]_{2}$, $[v\text{D}\text{D}Kv]_{1}$,
$[v\text{D}\text{D}Kv]_{2}$, $[vv\text{D}\text{D}K]$, $[\text{D}\text{D}Kvv]$,
$[vv\text{D}\text{D}Kvv]$\tabularnewline
\hline 
$\text{D}\text{D}a$ & $[v\text{D}\text{D}a]_{1}$, $[v\text{D}\text{D}a]_{2}$, $[\text{D}\text{D}av]$,
$[vv\text{D}\text{D}av]$\tabularnewline
\hline 
$\text{D}{}^{3}\!R$ & $[v\text{D}{}^{3}\!R]$, $[\text{D}{}^{3}\!Rv]$, $[v\text{D}{}^{3}\!Rvv]$\tabularnewline
\hline 
\end{tabular}
\par\end{centering}
\caption{Unfactorizable monomials an $d=3$: part I.}

\label{tab:mono3_1}
\end{table}

\begin{table}[H]
\renewcommand\arraystretch{1.5}
\begin{centering}
\begin{tabular}{|c|c|}
\hline 
B.b. & Unfactorizable monomials\tabularnewline
\hline 
$LLL$ & $[LLL]$, $[uLLLu]$\tabularnewline
\hline 
$LLb$ & $[uLLb]$\tabularnewline
\hline 
$Lbb$ & $[Lbb]$\tabularnewline
\hline 
$L^{3}\tilde{R}$ & $[L^{3}\tilde{R}]$, $[uL^{3}\tilde{R}u]$\tabularnewline
\hline 
$L\tilde{\text{D}}L$ & $[uL\tilde{\text{D}}L]_{1}$, $[uL\tilde{\text{D}}L]_{2}$, $[Lu\tilde{\text{D}}L]$,
$[L\tilde{\text{D}}Lu]$, $[uL\tilde{\text{D}}Luu]$, $[Lu\tilde{\text{D}}Luu]$\tabularnewline
\hline 
$L\tilde{\text{D}}b$ & $[uL\tilde{\text{D}}b]$, $[Lu\tilde{\text{D}}b]$, $[L\tilde{\text{D}}bu]$\tabularnewline
\hline 
$bbb$ & -\tabularnewline
\hline 
$b^{3}\tilde{R}$ & $[b^{3}\tilde{R}u]$\tabularnewline
\hline 
$b\tilde{\text{D}}b$ & $[bu\tilde{\text{D}}b]$, $[b\tilde{\text{D}}bu]$\tabularnewline
\hline 
$\tilde{\text{D}}\tilde{\text{D}}L$ & $[\tilde{\text{D}}\tilde{\text{D}}L]_{1}$, $[\tilde{\text{D}}\tilde{\text{D}}L]_{2}$,
$[u\tilde{\text{D}}\tilde{\text{D}}Lu]_{1}$, $[u\tilde{\text{D}}\tilde{\text{D}}Lu]_{2}$,
$[uu\tilde{\text{D}}\tilde{\text{D}}L]$, $[\tilde{\text{D}}\tilde{\text{D}}Luu]$,
$[uu\tilde{\text{D}}\tilde{\text{D}}Luu]$\tabularnewline
\hline 
$\tilde{\text{D}}\tilde{\text{D}}b$ & $[u\tilde{\text{D}}\tilde{\text{D}}b]_{1}$, $[u\tilde{\text{D}}\tilde{\text{D}}b]_{2}$,
$[\tilde{\text{D}}\tilde{\text{D}}bu]$, $[uu\tilde{\text{D}}\tilde{\text{D}}bu]$\tabularnewline
\hline 
$\tilde{\text{D}}{}^{3}\tilde{R}$ & $[u\tilde{\text{D}}{}^{3}\tilde{R}]$, $[\tilde{\text{D}}{}^{3}\tilde{R}u]$,
$[u\tilde{\text{D}}{}^{3}\tilde{R}uu]$\tabularnewline
\hline 
\end{tabular}
\par\end{centering}
\caption{Unfactorizable monomials an $d=3$, part II.}

\label{tab:mono3_2}
\end{table}

\begin{table}[H]
\renewcommand\arraystretch{1.5}
\begin{centering}
\begin{tabular}{|c|c|}
\hline 
B.b. & Unfactorizable monomials\tabularnewline
\hline 
$KKL$ & $[KKL]$, $[KvKvL]$, $[KKvLu]$\tabularnewline
\hline 
$KLL$ & $[KLL]$, $[KLuLu]$, $[KvLuL]$\tabularnewline
\hline 
$KKb$ & $[vKKb]$\tabularnewline
\hline 
$KLa$ & $[KvaLu]$, $[KvLua]$\tabularnewline
\hline 
$KLb$ & $[KvbLu]$, $[KvLub]$\tabularnewline
\hline 
$Kab$ & $[Kab]$\tabularnewline
\hline 
$Kbb$ & $[Kbb]$\tabularnewline
\hline 
$K^{3}\!\tilde{R}$ & $[K^{3}\!\tilde{R}]$, $[vK^{3}\!\tilde{R}u]$\tabularnewline
\hline 
$K\tilde{\text{D}}L$ & $[vK\tilde{\text{D}}L]$, $[Kv\tilde{\text{D}}L]$, $[Ku\tilde{\text{D}}L]$,
$[K\tilde{\text{D}}Lu]$, $[vK\tilde{\text{D}}Luu]$, $[Kv\tilde{\text{D}}Luu]$\tabularnewline
\hline 
$K\tilde{\text{D}}b$ & $[K\tilde{\text{D}}b]$, $[vKu\tilde{\text{D}}b]$, $[vK\tilde{\text{D}}bu]$\tabularnewline
\hline 
$LLa$ & $[uLLa]$\tabularnewline
\hline 
$Laa$ & $[Laa]$\tabularnewline
\hline 
$Lab$ & $[Lab]$\tabularnewline
\hline 
$L^{3}\!R$ & $[L^{3}\!R]$, $[uL^{3}\!Rv]$\tabularnewline
\hline 
$L\text{D}a$ & $[uL\text{D}a]$, $[Lv\text{D}a]$, $[L\text{D}av]$\tabularnewline
\hline 
$aab$ & -\tabularnewline
\hline 
$abb$ & -\tabularnewline
\hline 
$a^{3}\!\tilde{R}$ & $[a^{3}\!\tilde{R}u]$\tabularnewline
\hline 
$a\tilde{\text{D}}b$ & $[au\tilde{\text{D}}b]$, $[a\tilde{\text{D}}bu]$\tabularnewline
\hline 
$b^{3}\!R$ & $[b^{3}\!Rv]$\tabularnewline
\hline 
\end{tabular}
\par\end{centering}
\caption{Unfactorizable monomials an $d=3$, part III.}

\label{tab:mono3_3}
\end{table}

In the above tables, we define 
\begin{align}
[vK\text{D}K]_{1} & \coloneqq v^{a}K_{a}^{b}h^{cd}\text{D}_{b}K_{cd},\\{}
[vK\text{D}K]_{2} & \coloneqq v^{a}K_{a}^{b}h^{cd}\text{D}_{c}K_{bd},\\{}
[Kv\text{D}K] & \coloneqq K^{ab}v^{c}\text{D}_{c}K_{ab},\\{}
[K\text{D}Kv] & \coloneqq K^{ab}\text{D}_{a}K_{bc}v^{c},\\{}
[vK\text{D}Kvv] & \coloneqq v^{a}K_{a}^{b}\text{D}_{b}K_{cd}v^{c}v^{d},\\{}
[Kv\text{D}Kvv] & \coloneqq v^{a}K_{a}^{b}\text{D}_{c}K_{bd}v^{c}v^{d},
\end{align}
\begin{align}
[vKv\text{D}a] & \coloneqq v^{a}K_{a}^{b}v^{c}\text{D}_{c}a_{b},\\{}
[vK\text{D}av] & \coloneqq v^{a}K_{a}^{b}\text{D}_{b}a_{c}v^{c},
\end{align}

\begin{align}
[av\text{D}a] & \coloneqq a^{a}v^{b}\text{D}_{b}a_{a},\\{}
[a\text{D}av] & \coloneqq a^{a}\text{D}_{a}a_{b}v^{b},
\end{align}
\begin{align}
[\text{D}\text{D}K]_{1} & \coloneqq h^{ac}h^{bd}\text{D}_{a}\text{D}_{b}K_{cd},\\{}
[\text{D}\text{D}K]_{2} & \coloneqq h^{ab}h^{cd}\text{D}_{a}\text{D}_{b}K_{cd},\\{}
[v\text{D}\text{D}Kv]_{1} & \coloneqq v^{a}h^{bc}\text{D}_{a}\text{D}_{b}K_{cd}v^{d},\\{}
[v\text{D}\text{D}Kv]_{2} & \coloneqq v^{b}h^{ac}\text{D}_{a}\text{D}_{b}K_{cd}v^{d},\\{}
[vv\text{D}\text{D}K] & \coloneqq v^{a}v^{b}\text{D}_{a}\text{D}_{b}K_{cd}h^{cd},\\{}
[\text{D}\text{D}Kvv] & \coloneqq h^{ab}\text{D}_{a}\text{D}_{b}K_{cd}v^{c}v^{d},\\{}
[vv\text{D}\text{D}Kvv] & \coloneqq v^{a}v^{b}\text{D}_{a}\text{D}_{b}K_{cd}v^{c}v^{d},
\end{align}
\begin{align}
[v\text{D}\text{D}a]_{1} & \coloneqq v^{a}h^{bc}\text{D}_{a}\text{D}_{b}a_{c},\\{}
[v\text{D}\text{D}a]_{2} & \coloneqq v^{b}h^{ac}\text{D}_{a}\text{D}_{b}a_{c},\\{}
[\text{D}\text{D}av] & \coloneqq h^{ab}\text{D}_{a}\text{D}_{b}a_{c}v^{c},\\{}
[vv\text{D}\text{D}av] & \coloneqq v^{a}v^{b}\text{D}_{a}\text{D}_{b}a_{c}v^{c},
\end{align}
and
\begin{align}
[v\text{D}{}^{3}R] & \coloneqq v^{a}\text{D}_{a}{}^{3}R_{bc}h^{bc},\\{}
[\text{D}{}^{3}Rv] & \coloneqq h^{ab}\text{D}_{a}{}^{3}R_{bc}v^{c},\\{}
[v\text{D}{}^{3}Rvv] & \coloneqq v^{a}\text{D}_{a}{}^{3}R_{bc}v^{b}v^{c}.
\end{align}
Note that monomials $[\mathrm{D}\mathrm{D}K]_{1}$, $[\mathrm{D}\mathrm{D}K]_{2}$ in Table \ref{tab:mono3_1}, and $[\tilde{\mathrm{D}}\tilde{\mathrm{D}}L]_{1}$, $[\tilde{\mathrm{D}}\tilde{\mathrm{D}}L]_{2}$ in Table \ref{tab:mono3_2} are actually total derivatives at $d=3$. We deliberately keep them as they will contribute to factorizable monomials at higher orders.

\section{Correspondence to bi-scalar-tensor theory \label{sec:corr}}

As we have discussed above, the main motivation for studying the coupling of geometrical quantities of two foliations of hypersurfaces is to use it as a simpler and more straightforward method to construct ghost-free multiple scalar-tensor theories. In this section, it is thus interesting to see the explicit expressions of the corresponding bi-scalar-tensor theory.

To this end, we need to reverse the 3+1 decomposition procedure, i.e., to recover the normal vectors as gradients of the scalar fields and to rewrite Lie/spatial derivatives in terms of the spacetime covariant derivatives. This approach is usually referred to as the ``St\"{u}ckelberg trick'', which is used for covariant Ho\v{r}ava gravity \citep{Germani:2009yt,Blas:2009yd,Jacobson:2010mx,Blas:2010hb} and single-field spatially covariant gravity \citep{Gao:2020yzr,Hu:2021bbo,Hu:2024hzo}. After applying this procedure, the geometrical quantities of hypersurfaces are ``translated'' to the scalar-tensor expressions. We emphasize that expressions in both ``languages'', i.e., hypersurface and scalar-tensor theory, respect the full spacetime diffeomorphism.

All we need to do is replace the normal vectors $u_{a}$ and $v_{a}$ with the corresponding scalar-tensor expressions, i.e.,
\begin{equation}
u_{a}\rightarrow-\frac{\nabla_{a}\phi}{\sqrt{2X}},\quad v_{a}\rightarrow-\frac{\nabla_{a}\psi}{\sqrt{2Y}}.
\end{equation}
As a result, the induced metrics become
\begin{equation}
h_{ab}\rightarrow g_{ab}+\frac{\nabla_{a}\phi\nabla_{b}\phi}{2X},\quad l_{ab}\rightarrow g_{ab}+\frac{\nabla_{a}\psi\nabla_{b}\psi}{2Y}.
\end{equation}

The extrinsic curvature $K_{ab}$ and acceleration $a_{a}$ thus correspond to
\begin{equation}
K_{ab}\rightarrow h_{a}^{a'}h_{b}^{b'}\nabla_{(a'}u_{b')}=-\frac{1}{\sqrt{2X}}h_{a}^{a'}h_{b}^{b'}\nabla_{a'}\nabla_{b'}\phi,
\end{equation}
and
\begin{equation}
a_{a}\rightarrow h_{a}^{a'}u^{b}\nabla_{b}u_{a'}=\frac{1}{2X}h_{a}^{a'}\nabla^{b'}\phi\nabla_{a'}\nabla_{b'}\phi,
\end{equation}
respectively. We refer to \citep{Gao:2020yzr} for more explicit expressions.
The corresponding scalar-tensor expressions of $L_{ab}$ and $b_{a}$ can be
got by simply replacing $u_{a}$ by $v_{a}$. For example, the spatial
Ricci tensor of $\Sigma_{\psi}$ corresponds to
\begin{align}
^{3}\!\tilde{R}_{ab} & \rightarrow l_{a}^{a'}l_{b}^{b'}l^{cd}\left(R_{a'cb'd}-\nabla_{(a'}v_{b')}\nabla_{(c}v_{d)}+\nabla_{(a'}v_{d)}\nabla_{(c}v_{b')}\right)\nonumber \\
 & =l_{a}^{a'}l_{b}^{b'}l^{cd}\left[\,{}^{4}\!R_{a'cb'd}-\frac{1}{2Y}\left(\nabla_{a'}\nabla_{b'}\psi\nabla_{c}\nabla_{d}\psi-\nabla_{a'}\nabla_{d}\psi\nabla_{c}\nabla_{b'}\psi\right)\right].\label{Rictld_corr}
\end{align}
where $R_{a'cb'd}$ is the spacetime Riemann tensor. Expanding (\ref{Rictld_corr})
yields the explicit scalar-tensor expression.

We now focus on the scalar-tensor correspondence of the hypersurface monomials. For the monomials involving geometric quantities of $\Sigma_{\phi}$ (or $\Sigma_{\psi}$) only, the corresponding expressions describe a scalar-tensor theory for a single scalar field $\phi$ (or $\psi$), which has been given explicitly in \citep{Gao:2020yzr}. For example, for monomials of $\Sigma_{\phi}$,
\begin{equation}
K^{ab}K_{ab}\to\frac{1}{2X}\phi_{ab}\phi^{ab}+\frac{1}{2X^{2}}\phi^{a}\phi^{b}\phi_{a}^{\ c}\phi_{bc}+\frac{1}{8X^{3}}(\phi^{a}\phi^{b}\phi_{ab})^{2},
\end{equation}
\begin{equation}
K^{2}\to\frac{1}{2X}(\phi_{a}^{\ a})^{2}+\frac{1}{2X^{2}}(\phi^{a}\phi^{b}\phi_{ab})\phi_{c}^{\ c}+\frac{1}{8X^{3}}(\phi^{a}\phi^{b}\phi_{ab})^{2},
\end{equation}
\begin{equation}
a_{a}a^{a}\to\frac{1}{4X^{2}}\phi^{a}\phi^{b}\phi_{a}^{\ c}\phi_{bc}+\frac{1}{8X^{3}}(\phi^{a}\phi^{b}\phi_{ab})^{2},
\end{equation}
and
\begin{eqnarray}
^{3}\!R\equiv{}^{3}\!R_{ab}h^{ab} & \to & R+\frac{1}{2X}\Big(2R_{ab}\phi^{a}\phi^{b}+\phi_{ab}\phi^{ab}-(\phi^{a}{}_{a})^{2}\Big)\nonumber \\
 &  & +\frac{1}{2X^{2}}\Big(\phi^{a}\phi^{b}\phi_{a}{}^{c}\phi_{bc}-(\phi^{a}\phi^{b}\phi_{ab})\phi^{c}{}_{c}\Big),
\end{eqnarray}
where we denote the factorized terms with brackets and $\phi_{a}\coloneqq\nabla_{a}\phi$,
$\phi_{ba}\coloneqq\nabla_{a}\nabla_{b}\phi$. The scalar-tensor correspondence
of monomials of $\Sigma_{\psi}$ can be read simply by replacing $\phi$
by $\psi$ in the above, e.g.,
\begin{equation}
L^{ab}L_{ab}\equiv L_{ab}L_{cd}l^{ac}l^{bd}\rightarrow\frac{1}{2Y}\psi_{ab}\psi^{ab}+\frac{1}{2Y^{2}}\psi^{a}\psi^{b}\psi_{a}^{\ c}\psi_{bc}+\frac{1}{8Y^{3}}(\psi^{a}\psi^{b}\psi_{ab})^{2}.
\end{equation}

For our purposes, it is interesting to derive the scalar-tensor expressions corresponding to the ``mixing'' terms, i.e., monomials constructed from quantities belonging to both $\Sigma_{\phi}$ and $\Sigma_{\psi}$, which inevitably induce couplings between the two scalar fields $\phi$ and $\psi$. For example, the term $a_{a} v^{a}$, which does not exist in the single-field case, corresponds to
\begin{equation}
a_{a}v^{a}\rightarrow-\frac{1}{2X\sqrt{2Y}}\left(\psi^{a}-\frac{Z}{X}\phi^{a}\right)\phi^{b}\phi_{ab},
\end{equation}
where the canonical kinetic terms $X$, $Y$ and the mixing kinetic
term are given in (\ref{XY_def}) and (\ref{Z_def}), respectively.
We also have
\begin{equation}
K_{ab}v^{a}v^{b}\rightarrow-\frac{1}{\sqrt{2X}2Y}\phi_{ab}\left(\psi^{a}-\frac{Z}{X}\phi^{a}\right)\left(\psi^{b}-\frac{Z}{X}\phi^{a}\right).
\end{equation}
Clearly, such kind of terms does not exist in the case of a single
scalar field.

In the remainder of this work, we consider a polynomial-type Lagrangian for illustration, in which the monomials are of $d=2$ but without mixings through $u_{a}$ and $v_{a}$\footnote{That is, terms involving contractions with $u_{a}$ and $v_{a}$, such as $\left[KvKv\right]$, $\left[LuLu\right]$, $\left[Kva\right]$, etc., in Table \ref{tab:mono2}, are not considered.}. The action is given by
\begin{equation}
S_{2}=\int\mathrm{d}^{4}x\sqrt{-g}\mathcal{L}_{2},
\end{equation}
with the Lagrangian
\begin{eqnarray}
\mathcal{L}_{2} & = & c_{1}K_{ab}K^{ab}+c_{2}K^{2}+c_{3}K_{ab}L^{ab}+c_{4}KL+c_{5}L_{ab}L^{ab}+c_{6}L^{2}\nonumber \\
 &  & +c_{7}a_{a}a^{a}+c_{8}a_{a}b^{a}+c_{9}b_{a}b^{a}+c_{10}{}^{3}\!R+c_{11}{}^{3}\!\tilde{R},\label{Lagd2}
\end{eqnarray}
where the coefficients $c_{i}$, $i=1,\ldots,11$, are general functions of $N$, $M$, $\alpha$, $\phi$, and $\psi$. Keep in mind that all the indices are raised and lowered by the spacetime metric or the ``adapted'' spatial metric, e.g., $K_{ab}K^{ab} = K_{ab}K_{cd}g^{ac}g^{bd} = K_{ab}K_{cd}h^{ac}h^{bd}$. Note that in the single-field case, the determinant of the spacetime metric is usually written explicitly in terms of $\sqrt{-g} = N\sqrt{h}$. Here, since we have two foliations of hypersurfaces and there is no natural decomposition with respect to one of them, we keep the determinant of the spacetime metric as $\sqrt{-g}$.

For the monomials in (\ref{Lagd2}), after some manipulations, we
find
\begin{eqnarray}
K_{ab}L^{ab} & \to & \frac{1}{2X^{1/2}Y^{1/2}}\phi^{ab}\psi_{ab}+\frac{1}{2X^{3/2}Y^{1/2}}\phi^{a}\phi^{b}\phi_{a}{}^{c}\psi_{bc}+\frac{1}{2X^{1/2}Y^{3/2}}\phi_{a}{}^{c}\psi^{a}\psi^{b}\psi_{bc}\nonumber \\
 &  & +\frac{1}{8X^{5/2}Y^{1/2}}(\phi^{a}\phi^{b}\phi_{ab})(\phi^{c}\phi^{d}\psi_{cd})+\frac{1}{8X^{1/2}Y^{5/2}}(\phi_{ab}\psi^{a}\psi^{b})(\psi^{d}\psi^{c}\psi_{dc})\nonumber \\
 &  & +\frac{1}{4X^{3/2}Y^{3/2}}(\phi^{a}\psi^{b}\psi_{ab})(\phi^{d}\phi_{dc}\psi^{c})-\frac{Z}{2X^{3/2}Y^{3/2}}\phi^{a}\phi_{a}{}^{c}\psi^{b}\psi_{bc}\nonumber \\
 &  & -\frac{Z}{4X^{5/2}Y^{3/2}}(\phi^{a}\phi^{b}\phi_{ab})(\phi^{d}\psi^{c}\psi_{dc})-\frac{Z}{4X^{3/2}Y^{5/2}}(\phi^{a}\phi_{ab}\psi^{b})(\psi^{d}\psi^{c}\psi_{dc})\nonumber \\
 &  & +\frac{Z^{2}}{8X^{5/2}Y^{5/2}}(\phi^{a}\phi^{b}\phi_{ab})(\psi^{d}\psi^{c}\psi_{dc}),
\end{eqnarray}
\begin{eqnarray}
KL & \to & \frac{(\psi^{a}{}_{a})}{2X^{1/2}Y^{1/2}}(\phi^{a}{}_{a})^{2}+\frac{1}{4X^{1/2}Y^{3/2}}(\psi^{a}\psi^{b}\psi_{ab})\phi^{c}{}_{c}\\
 &  & +\frac{1}{4X^{3/2}Y^{1/2}}(\phi^{a}\phi^{b}\phi_{ab})\psi^{c}{}_{c}+\frac{1}{8X^{3/2}Y^{3/2}}(\phi^{a}\phi^{b}\phi_{ab})(\psi^{c}\psi^{d}\psi_{cd}),
\end{eqnarray}
\begin{eqnarray}
K_{ab}l^{ab} & \to & -\frac{1}{\sqrt{2X}}\phi^{a}{}_{a}-\frac{1}{2\sqrt{2}X^{3/2}}\phi^{a}\phi^{b}\phi_{ab}-\frac{1}{2\sqrt{2}X^{1/2}Y}\phi_{ab}\psi^{a}\psi^{b}\nonumber \\
 &  & +\frac{Z}{\sqrt{2}X^{3/2}Y}\phi^{a}\phi_{ab}\psi^{b}-\frac{Z^{2}}{2\sqrt{2X^{5}}Y}\phi^{a}\phi^{b}\phi_{ab},
\end{eqnarray}
and
\begin{eqnarray}
^{3}\!R_{ab}l^{ab} & \to & R+\frac{1}{2X}\Big(2R_{ab}\phi^{a}\phi^{b}+\phi_{ab}\phi^{ab}-(\phi^{a}{}_{a})^{2}\Big)+\frac{1}{2Y}R_{ab}\psi^{a}\psi^{b}\nonumber \\
 &  & +\frac{1}{4XY}\Big(R_{acbd}\phi^{a}\phi^{b}\psi^{c}\psi^{d}+\phi_{a}{}^{c}\phi_{bc}\psi^{a}\psi^{b}-(\phi_{ab}\psi^{a}\psi^{b})\phi^{c}{}_{c}\Big)-\frac{Z}{XY}R_{ab}\phi^{a}\psi^{b}\nonumber \\
 &  & +\frac{1}{2X^{2}}\Big(\phi^{a}\phi^{b}\phi_{a}{}^{c}\phi_{bc}-(\phi^{a}\phi^{b}\phi_{ab})\phi^{c}{}_{c}\Big)\nonumber \\
 &  & +\frac{1}{8X^{2}Y}\Big((\psi^{a}\phi^{b}\phi_{ba})^{2}-(\phi^{a}\phi^{b}\phi_{ab})(\phi_{dc}\psi^{d}\psi^{c})\Big)\nonumber \\
 &  & +\frac{Z}{2X^{2}Y}\Big((\phi^{a}\phi_{ab}\psi^{b})\phi^{c}{}_{c}-\phi^{a}\phi_{a}{}^{c}\phi_{bc}\psi^{b}\Big)+\frac{Z^{2}}{2X^{2}Y}R_{ab}\phi^{a}\phi^{b}\nonumber \\
 &  & +\frac{Z^{2}}{4X^{3}Y}\Big(\phi^{a}\phi^{b}\phi_{a}{}^{c}\phi_{bc}-(\phi^{a}\phi^{b}\phi_{ab})\phi^{c}{}_{c}\Big).
\end{eqnarray}

\section{Perturbation analysis \label{sec:pert}}

After applying the covariant correspondence, the Lagrangian (\ref{Lagd2}) corresponds to a generally covariant bi-scalar-tensor theory, in which hypersurface monomials are replaced by their generally covariant counterparts. The coefficients $c_{i}$ become general functions of $X$, $Y$, $Z$, $\phi$, and $\psi$, which we denote as $f_{i} = f_{i}(X,Y,Z,\phi,\psi)$. 

The resulting Lagrangian explicitly includes second-order derivatives of the scalar fields and mixing derivative terms of different scalar fields. Due to the presence of higher-order time derivatives, one may expect that there are 6 degrees of freedom, i.e., 4 scalar modes and 2 tensor modes, propagating in general. However, with our construction, there are, in fact, only two dynamical scalar modes, while another two scalar modes become the so-called instantaneous or shadowy modes. In \citep{DeFelice:2021hps}, it is argued that the non-linear shadowy mode could be completely determined once an appropriate boundary condition is imposed on a spatial boundary, which means it does not propagate, in fact. 

In the following, we will examine whether the theory (\ref{Lagd2}) propagates only two scalar modes or not through a cosmological perturbation analysis. Since the Lagrangian respects full general covariance, we are allowed to fix the unitary gauge as in the case of a single scalar field. To be precise, one of the two scalar fields is chosen to specify the hypersurfaces that are employed to make the 3+1 decomposition. Since the Lagrangian is completely symmetric with respect to $\phi$ and $\psi$, without loss of generality, we fix the unitary gauge and the associated coordinate system by choosing $\phi = \phi\left(t\right)$ and thus $\delta\phi = 0$ in the perturbation language. 

After fixing the unitary gauge, the spacetime metric is given in terms of the ADM variables as usual,
\begin{equation}
g_{\mu\nu}=\begin{pmatrix}-N^{2}+N^{k}N_{k} & N_{i}\\
N_{j} & h_{ij}
\end{pmatrix},
\end{equation}
where the index of the shift vector is raised by the spatial metric $h_{ij}$, i.e., $N^{i} \equiv h^{ij} N_{j}$. Around a Friedmann-Robertson-Walker (FRW) background, the perturbed ADM variables are parametrized by
\begin{eqnarray}
N & = & e^{A},\\
N_{i} & = & a(\partial_{i}B+S_{i}),\\
h_{ij} & = & a^{2}\delta_{ik}(e^{\bm{H}})_{\ j}^{k},
\end{eqnarray}
where the FRW background corresponds to $N=1$, $N_{i}=0$, $h_{ij}=a^{2}\delta_{ij}$.
$H_{ij}$ can be further decomposed as
\begin{equation}
H_{ij}=2\zeta\delta_{ij}+(\partial_{i}\partial_{j}-\frac{1}{3}\delta_{ij}\partial^{2})E+\partial_{i}F_{j}+\partial_{j}F_{i}+\gamma_{ij},
\end{equation}
with constraints
\begin{equation}
0=\partial_{i}F^{i}=\partial_{i}S^{i}=\partial_{i}\gamma_{\phantom{i}j}^{i}=\gamma_{\phantom{i}i}^{i}.
\end{equation}
We thus have 4 scalar modes $A$, $B$, $E$ and $\zeta$, 4 vector modes
$S_{i}$ and $F_{i}$, as well as 2 tensor modes $\gamma_{ij}$.

The Lagrangian can be expanded up to quadratic order in perturbations as
\begin{equation}
L_{2}\equiv N\sqrt{h}\mathcal{L}_{2}^{\text{GST}}\simeq L^{0}+L^{1}+L^{2\text{s}}+L^{2\text{v}}+L^{2\text{t}}+\cdots,
\end{equation}
where the upper index is a notation indicating the order of perturbations and ``$\cdots$'' stands for the higher orders of perturbations. At the quadratic order, the scalar, vector, and tensor modes are decoupled around a homogeneous and isotropic background, which allows us to use ``s, v, t'' to denote different sectors of perturbations.

\subsection{Background equations of motion}

The background value of the Lagrangian is a function of time
\begin{equation}
L^{0}=3\mathcal{F}_{1}a^{3}H^{2},
\end{equation}
where 
\begin{equation}
\mathcal{F}_{1}\coloneqq f_{1}+3f_{2}+f_{3}+3f_{4}+f_{5}+3f_{6},\label{calF}
\end{equation}
and the coefficients $f_{i}$ should be evaluated at their background values. After some manipulations, we obtain the first-order Lagrangian for the perturbations, which is a linear combination of $A, \zeta, \delta\psi, \delta\phi$,
\begin{eqnarray}
L^{1} & \simeq & -3a^{3}\Big(2\dot{\mathcal{F}_{1}}H+H^{2}\mathcal{F}_{1}+2\mathcal{F}_{1}\frac{\ddot{a}}{a}\Big)\zeta\nonumber \\
 &  & -3a^{3}H^{2}\Big(\mathcal{F}_{1}+\mathcal{F}_{1X}\dot{\bar{\phi}}{}^{2}+\mathcal{F}_{1Y}\dot{\bar{\psi}}{}^{2}+\mathcal{F}_{1Z}\dot{\bar{\phi}}\dot{\bar{\psi}}\Big)A\nonumber \\
 &  & +3a^{3}\Big(\mathcal{F}_{1\psi}H^{2}-\mathcal{F}_{1Y}H^{3}\dot{\bar{\psi}}-\frac{1}{2}\mathcal{F}_{1Z}H^{3}\dot{\bar{\phi}}-\mathcal{\dot{F}}_{1Y}H^{2}\dot{\bar{\psi}}-\frac{1}{2}\mathcal{\dot{F}}_{1Z}H^{2}\dot{\bar{\phi}}\nonumber \\
 &  & \quad-\mathcal{F}_{1Y}H^{2}\ddot{\bar{\psi}}-\frac{1}{2}\mathcal{F}_{1Z}H^{2}\ddot{\bar{\phi}}-2\mathcal{F}_{1Y}H\dot{\bar{\psi}}\frac{\ddot{a}}{a}-\mathcal{F}_{1Z}H\dot{\bar{\phi}}\frac{\ddot{a}}{a}\Big)\delta\psi\nonumber \\
 &  & +3a^{3}\Big(\mathcal{F}_{1\phi}H^{2}-\mathcal{F}_{1X}H^{3}\dot{\bar{\phi}}-\frac{1}{2}\mathcal{F}_{1Z}H^{3}\dot{\bar{\psi}}-\mathcal{\dot{F}}_{1X}H^{2}\dot{\bar{\phi}}-\frac{1}{2}\mathcal{\dot{F}}_{1Z}H^{2}\dot{\bar{\psi}}\nonumber \\
 &  & \quad-\mathcal{F}_{1X}H^{2}\ddot{\bar{\phi}}-\frac{1}{2}\mathcal{F}_{1Z}H^{2}\ddot{\bar{\psi}}-2\mathcal{F}_{1X}H\dot{\bar{\phi}}\frac{\ddot{a}}{a}-\mathcal{F}_{1Z}H\dot{\bar{\psi}}\frac{\ddot{a}}{a}\Big)\delta\phi.\label{Lag1}
\end{eqnarray}
Note that at this point, we have not yet fixed the unitary gauge with $\delta\phi = 0$.

The vanishing of the first-order Lagrangian yields the background equations of motion, namely the Friedmann equations.
\begin{eqnarray}
\mathcal{F}_{1}+\mathcal{F}_{1X}\dot{\bar{\phi}}^{2}+\mathcal{F}_{1Z}\dot{\bar{\phi}}\dot{\bar{\psi}}+\mathcal{F}_{1Y}\dot{\bar{\psi}}^{2} & = & 0,\label{bgeom_A}\\
2H\dot{\mathcal{F}}_{1}+H^{2}\mathcal{F}_{1}+2\frac{\ddot{a}}{a}\mathcal{F}_{1} & = & 0,\label{bgeom_zeta}
\end{eqnarray}
and the continuity equations 
\begin{eqnarray}
H\mathcal{F}_{1\phi} & = & H\partial_{t}(\mathcal{F}_{1X}\dot{\bar{\phi}})+H^{2}\mathcal{F}_{1X}\dot{\bar{\phi}}+2\frac{\ddot{a}}{a}\mathcal{F}_{1X}\dot{\bar{\phi}}\nonumber \\
 &  & +\frac{1}{2}H\partial_{t}(\mathcal{F}_{1Z}\dot{\bar{\psi}})+\frac{1}{2}\mathcal{F}_{1Z}H^{2}\dot{\bar{\psi}}+\mathcal{F}_{1Z}\dot{\bar{\psi}}\frac{\ddot{a}}{a},\label{bgeom_phi}\\
H\mathcal{F}_{1\psi} & = & H\partial_{t}(\mathcal{F}_{1Y}\dot{\bar{\psi}})+H^{2}\mathcal{F}_{1Y}\dot{\bar{\psi}}+2\frac{\ddot{a}}{a}\mathcal{F}_{1Y}\dot{\bar{\psi}}\nonumber \\
 &  & +\frac{1}{2}H\partial_{t}(\mathcal{F}_{1Z}\dot{\bar{\phi}})+\frac{1}{2}H^{2}\mathcal{F}_{1Z}\dot{\bar{\phi}}+\frac{\ddot{a}}{a}\mathcal{F}_{1Z}\dot{\bar{\phi}},\label{bgeom_psi}
\end{eqnarray}
which are also the background equations of motion for the scalar fields.

The continuity equations (\ref{bgeom_phi}) and (\ref{bgeom_psi})
can be equivalently written as
\begin{eqnarray}
H^{2}\mathcal{F}_{1\phi} & = & \frac{1}{a^{3}}\partial_{t}\bigg[a^{3}H^{2}\left(\mathcal{F}_{1X}\dot{\bar{\phi}}+\frac{1}{2}\mathcal{F}_{1Z}\dot{\bar{\psi}}\right)\bigg],\\
H^{2}\mathcal{F}_{1\psi} & = & \frac{1}{a^{3}}\partial_{t}\bigg[a^{3}H^{2}\left(\mathcal{F}_{1Y}\dot{\bar{\psi}}+\frac{1}{2}\mathcal{F}_{1Z}\dot{\bar{\phi}}\right)\bigg].
\end{eqnarray}
After combining the two equations (\ref{bgeom_A}) and (\ref{bgeom_zeta}),
we have
\begin{eqnarray}
H\mathcal{F}_{1\phi}\dot{\bar{\phi}}+H\mathcal{F}_{1\psi}\dot{\bar{\psi}} & = & H\partial_{t}\Big(\mathcal{F}_{1X}\dot{\bar{\phi}}^{2}\Big)-H\mathcal{F}_{1X}\dot{\bar{\phi}}\ddot{\bar{\phi}}+H^{2}\mathcal{F}_{1X}\dot{\bar{\phi}}^{2}\nonumber \\
 &  & +2\frac{\ddot{a}}{a}\mathcal{F}_{1X}\dot{\bar{\phi}}^{2}+H\partial_{t}\Big(\mathcal{F}_{1Y}\dot{\bar{\psi}}^{2}\Big)-H\mathcal{F}_{1Y}\dot{\bar{\psi}}\ddot{\bar{\psi}}\nonumber \\
 &  & +H^{2}\mathcal{F}_{1Y}\dot{\bar{\psi}}^{2}+2\frac{\ddot{a}}{a}\mathcal{F}_{1Y}\dot{\bar{\psi}}^{2}+H\partial_{t}\Big(\mathcal{F}_{1Z}\dot{\bar{\phi}}\dot{\bar{\psi}}\Big)\nonumber \\
 &  & -\frac{1}{2}H\mathcal{F}_{1Z}\dot{\bar{\psi}}\ddot{\bar{\phi}}-\frac{1}{2}H\mathcal{F}_{1Z}\dot{\bar{\phi}}\ddot{\bar{\psi}}\nonumber \\
 &  & +H^{2}\mathcal{F}_{1Z}\dot{\bar{\phi}}\dot{\bar{\psi}}+2\frac{\ddot{a}}{a}\mathcal{F}_{1Z}\dot{\bar{\phi}}\dot{\bar{\psi}},
\end{eqnarray}
which is a linear combination of (\ref{bgeom_phi}) and (\ref{bgeom_psi}). This means that one of these four equations (\ref{bgeom_A})-(\ref{bgeom_psi}) is actually not independent, which is a result of general covariance. This also justifies that we are free to choose one of the two scalar fields to fix the unitary gauge, and no information is lost for the background equations of motion.

\subsection{The linear scalar perturbations}

As we mentioned before, after making the covariant correspondence, the general covariance of the resulting bi-scalar-tensor theory becomes manifest. We are free to fix the coordinates to simplify the calculations. Without loss of generality, we choose the uniform $\phi$ gauge with $E = \delta\phi = 0$. After utilizing the background equations of motion, the quadratic Lagrangian for the scalar perturbations $A$, $B$, $\zeta$, and $\delta\psi$ takes the form
\begin{eqnarray}
L^{\text{2s}} & \simeq & \mathcal{C}_{\dot{\zeta}\dot{\zeta}}\dot{\zeta}^{2}+\mathcal{C}_{\dot{\zeta}\delta\dot{\psi}}\dot{\zeta}\delta\dot{\psi}+\mathcal{C}_{\dot{\zeta}\partial^{2}B}\dot{\zeta}\partial^{2}B+\mathcal{C}_{\delta\dot{\psi}\delta\dot{\psi}}\delta\dot{\psi}^{2}+\mathcal{C}_{\partial\delta\dot{\psi}\partial\delta\dot{\psi}}\partial_{i}\delta\dot{\psi}\partial^{i}\delta\dot{\psi}\nonumber \\
 &  & +\mathcal{C}_{\delta\dot{\psi}A}\delta\dot{\psi}A+\mathcal{C}_{\partial\delta\dot{\psi}\partial A}\partial^{i}\delta\dot{\psi}\partial_{i}A+\mathcal{C}_{\delta\dot{\psi}\partial^{2}B}\delta\dot{\psi}\partial^{2}B+\mathcal{C}_{AA}A^{2}\nonumber \\
 &  & +\mathcal{C}_{\partial A\partial A}\partial^{i}A\partial_{i}A+\mathcal{C}_{\partial^{2}B\partial^{2}B}\partial^{2}B\partial^{2}B\nonumber \\
 &  & +\mathcal{C}_{\partial\zeta\partial\zeta}\partial^{i}\zeta\partial_{i}\zeta+\mathcal{C}_{\zeta\delta\dot{\psi}}\zeta\delta\dot{\psi}+\mathcal{C}_{\partial^{2}\zeta\delta\dot{\psi}}\partial^{2}\zeta\delta\dot{\psi}+\mathcal{C}_{\zeta\delta\psi}\zeta\delta\psi+\mathcal{C}_{\zeta\partial^{2}\delta\psi}\zeta\partial^{2}\delta\psi\nonumber \\
 &  & +\mathcal{C}_{\partial^{2}\zeta A}\partial^{2}\zeta A+\mathcal{C}_{\delta\psi\delta\psi}\delta\psi^{2}+\mathcal{C}_{\partial\delta\psi\partial\delta\psi}\partial^{i}\delta\psi\partial_{i}\delta\psi+\mathcal{C}_{\partial^{2}\delta\psi\partial^{2}\delta\psi}\partial^{2}\delta\psi\partial^{2}\delta\psi\nonumber \\
 &  & +\mathcal{C}_{\delta\psi A}\delta\psi A+\mathcal{C}_{\partial^{2}\delta\psi A}\partial^{2}\delta\psi A+\mathcal{C}_{\delta\psi\partial^{2}B}\delta\psi\partial^{2}B+\mathcal{C}_{\partial^{2}\delta\psi\partial^{2}B}\partial^{2}\delta\psi\partial^{2}B.\label{Lag2s}
\end{eqnarray}
In (\ref{Lag2s}), $\mathcal{C}_{\dot{\zeta}\dot{\zeta}}$, $\mathcal{C}_{\dot{\zeta}\delta\dot{\psi}}$, etc., are coefficients depending on the background quantities, whose explicit expressions are listed in Appendix \ref{app:coeff}.

According to (\ref{Lag2s}), there are no time derivatives acting on $A$ and $B$, which implies that they play the role of auxiliary variables. On the other hand, $\zeta$ and $\delta\psi$ acquire time derivatives explicitly, making them dynamical variables. We can solve for them and plug the solutions into the Lagrangian (\ref{Lag2s}) to obtain the effective Lagrangian for the dynamical variables $\zeta$ and $\delta\psi$. Varying the action with respect to $\partial^{2}B$ yields the momentum constraint, from which we can solve for $B$ in terms of $\zeta$ and $\delta\psi$ as
\begin{eqnarray}
	\partial^{2}B & = & a\frac{\mathcal{F}_{1}}{\mathcal{F}_{2}}\dot{\zeta}+\frac{aH}{\mathcal{F}_{2}}\Big(\frac{\mathcal{F}_{1Z}}{2}\dot{\bar{\phi}}+\mathcal{F}_{1Y}\dot{\bar{\psi}}\Big)\delta\dot{\psi}\nonumber \\
	&  & +\frac{aH}{\mathcal{F}_{2}}\Big(\mathcal{F}_{1\psi}-\frac{3}{4}H\mathcal{F}_{1Z}\dot{\bar{\phi}}-\frac{3}{2}H\mathcal{F}_{1Y}\dot{\bar{\psi}}\Big)\delta\psi\nonumber \\
	&  & -\frac{f_{3}+f_{4}+2f_{5}+2f_{6}}{2a\mathcal{F}_{2}\dot{\bar{\psi}}}\partial^{2}\delta\psi,
\end{eqnarray}
with $\mathcal{F}_{2} \coloneqq f_{1} + f_{2} + f_{3} + f_{4} + f_{5} + f_{6} \neq 0$. After plugging the solution for $B$ into (\ref{Lag2s}), we can vary it with respect to \textbf{$A$}. The solution for $A$ is proportional to $\left(1 - \frac{f_{7} + f_{8} + f_{9}}{\mathcal{F}_{4} H^{2}} \frac{\partial^{2}}{a^{2}}\right)^{-1}$. In the following, we assume $f_{7} + f_{8} + f_{9} = 0$\footnote{If $f_{7} + f_{8} + f_{9} \neq 0$, the resulting effective Lagrangian for $\zeta$ will have nonlocal features since spatial derivatives will arise in the determinant.}. For the sake of simplicity, we split the solution for $A$ into two parts,
\begin{equation}
A=A_{\mathrm{k}}+\tilde{A},
\end{equation}
with
\begin{eqnarray}
A_{\mathrm{k}} & = & \frac{3}{4\mathcal{F}_{4}}\Big(3\mathcal{F}_{1Z}\dot{\bar{\phi}}+\mathcal{F}_{1XZ}\dot{\bar{\phi}}^{3}+6\mathcal{F}_{1Y}\dot{\bar{\psi}}+\mathcal{F}_{1ZZ}\dot{\bar{\phi}}^{2}\dot{\bar{\psi}}+2\mathcal{F}_{1XY}\dot{\bar{\phi}}^{2}\dot{\bar{\psi}}\nonumber \\
 &  & \quad+3\mathcal{F}_{1YZ}\dot{\bar{\phi}}\dot{\bar{\psi}}^{2}+2\mathcal{F}_{1YY}\dot{\bar{\psi}}^{3}\Big)\delta\dot{\psi}-\frac{f_{8}+2f_{9}}{2\mathcal{F}_{4}H^{2}\dot{\bar{\psi}}}\frac{\partial^{2}}{a^{2}}\delta\dot{\psi},
\end{eqnarray}
which will contribute to the kinetic term of $\delta\psi$, and
\begin{eqnarray}
\tilde{A} & = & \frac{3}{2\mathcal{F}_{4}}\Big(\mathcal{F}_{1\psi}+\mathcal{F}_{1\psi X}\dot{\bar{\phi}}^{2}+\mathcal{F}_{1\psi Z}\dot{\bar{\phi}}\dot{\bar{\psi}}+\mathcal{F}_{1\psi Y}\dot{\bar{\psi}}^{2}\Big)\delta\psi\nonumber \\
 &  & +\frac{1}{2H\mathcal{F}_{4}}\Big[4f_{11Z}\dot{\bar{\phi}}-\mathcal{F}_{3Z}\dot{\bar{\phi}}+\frac{1}{\dot{\bar{\psi}}}\Big(-4f_{11}-\mathcal{F}_{3}+4f_{11X}\dot{\bar{\phi}}^{2}\nonumber \\
 &  & \quad-\mathcal{F}_{3X}\dot{\bar{\phi}}^{2}+4f_{11Y}\dot{\bar{\psi}}^{2}-\mathcal{F}_{3Y}\dot{\bar{\psi}}^{2}+\frac{f_{8}\ddot{\bar{\psi}}}{H\dot{\bar{\psi}}}+\frac{2f_{9}\ddot{\bar{\psi}}}{H\dot{\bar{\psi}}}\Big)\Big]\frac{\partial^{2}}{a^{2}}\delta\psi\nonumber \\
 &  & +\frac{2}{\mathcal{F}_{4}H^{2}}\Big(f_{10}+2f_{11}-f_{10X}\dot{\bar{\phi}}^{2}-f_{11X}\dot{\bar{\phi}}^{2}-f_{10Z}\dot{\bar{\phi}}\dot{\bar{\psi}}-f_{11Z}\dot{\bar{\phi}}\dot{\bar{\psi}}\nonumber \\
 &  & \quad-f_{10Y}\dot{\bar{\psi}}^{2}-f_{11Y}\dot{\bar{\psi}}^{2}\Big)\frac{\partial^{2}}{a^{2}}\zeta,
\end{eqnarray}
which is irrelevant to the kinetic terms of $\zeta$ and $\delta\psi$.
In the above, the non-vanishing coefficient $\mathcal{F}_{4}$ is
defined by
\begin{eqnarray}
\mathcal{F}_{4} & \coloneqq & \frac{3}{2}\mathcal{F}_{1}+6\mathcal{F}_{1X}\dot{\bar{\phi}}^{2}+\frac{3}{2}\mathcal{F}_{1XX}\dot{\bar{\phi}}^{4}+6\mathcal{F}_{1Z}\dot{\bar{\phi}}\dot{\bar{\psi}}+3\mathcal{F}_{1XZ}\dot{\bar{\phi}}^{3}\dot{\bar{\psi}}\nonumber \\
 &  & +6\mathcal{F}_{1Y}\dot{\bar{\psi}}^{2}+\frac{3}{2}\mathcal{F}_{1ZZ}\dot{\bar{\phi}}^{2}\dot{\bar{\psi}}^{2}+3\mathcal{F}_{1XY}\dot{\bar{\phi}}^{2}\dot{\bar{\psi}}^{2}\nonumber \\
 &  & +3\mathcal{F}_{1YZ}\dot{\bar{\phi}}\dot{\bar{\psi}}^{3}+\tfrac{3}{2}\mathcal{F}_{1YY}\dot{\bar{\psi}}^{4}.
\end{eqnarray}

Finally, substituting the solution of $A$ into the original Lagrangian
yields the effective Lagrangian for $\zeta$ and $\delta\psi$,
\begin{eqnarray}
L^{\text{2s}} & = & \hat{\mathcal{C}}_{\dot{\zeta}\dot{\zeta}}\dot{\zeta}^{2}+\hat{\mathcal{C}}_{\dot{\zeta}\delta\dot{\psi}}\dot{\zeta}\delta\dot{\psi}+\hat{\mathcal{C}}_{\delta\dot{\psi}\delta\dot{\psi}}\delta\dot{\psi}^{2}+\hat{\mathcal{C}}_{\partial\delta\dot{\psi}\partial\delta\dot{\psi}}\partial_{i}\delta\dot{\psi}\partial^{i}\delta\dot{\psi}+\hat{\mathcal{C}}_{\partial^{2}\delta\dot{\psi}\partial^{2}\delta\dot{\psi}}(\partial^{2}\delta\dot{\psi})^{2}\nonumber \\
 &  & +\hat{\mathcal{C}}_{\zeta\delta\dot{\psi}}\zeta\delta\dot{\psi}+\hat{\mathcal{C}}_{\partial^{2}\zeta\delta\dot{\psi}}\partial^{2}\zeta\delta\dot{\psi}+\hat{\mathcal{C}}_{\partial^{2}\zeta\partial^{2}\delta\dot{\psi}}\partial^{2}\zeta\partial^{2}\delta\dot{\psi}\nonumber \\
 &  & +\hat{\mathcal{C}}_{\partial\zeta\partial\zeta}\partial^{i}\zeta\partial_{i}\zeta+\hat{\mathcal{C}}_{\partial^{2}\zeta\partial^{2}\zeta}(\partial^{2}\zeta)^{2}+\hat{\mathcal{C}}_{\zeta\delta\psi}\zeta\delta\psi+\hat{\mathcal{C}}_{\zeta\partial^{2}\delta\psi}\zeta\partial^{2}\delta\psi+\hat{\mathcal{C}}_{\partial^{2}\zeta\partial^{2}\delta\psi}\partial^{2}\zeta\partial^{2}\delta\psi\nonumber \\
 &  & +\hat{\mathcal{C}}_{\delta\psi\delta\psi}(\delta\psi)^{2}+\hat{\mathcal{C}}_{\partial\delta\psi\partial\delta\psi}\partial^{i}\delta\psi\partial_{i}\delta\psi+\hat{\mathcal{C}}_{\partial^{2}\delta\psi\partial^{2}\delta\psi}(\partial^{2}\delta\psi)^{2},\label{Lag2sfin}
\end{eqnarray}
where the coefficients $\hat{\mathcal{C}}_{\dot{\zeta}\dot{\zeta}}$,
$\hat{\mathcal{C}}_{\dot{\zeta}\delta\dot{\psi}}$ etc. are given
in Appendix \ref{app:coeff}. Generally, the Hessian matrix of the kinetic
terms $\dot{\zeta}$ and $\delta\dot{\psi}$ 
\begin{equation}
\bm{H}\coloneqq\left(\begin{array}{cc}
\hat{\mathcal{C}}_{\dot{\zeta}\dot{\zeta}} & \frac{1}{2}\hat{\mathcal{C}}_{\dot{\zeta}\delta\dot{\psi}}\\
\frac{1}{2}\hat{\mathcal{C}}_{\dot{\zeta}\delta\dot{\psi}} & \hat{\mathcal{C}}_{\delta\dot{\psi}\delta\dot{\psi}}-\hat{\mathcal{C}}_{\partial\delta\dot{\psi}\partial\delta\dot{\psi}}\partial^{2}+\hat{\mathcal{C}}_{\partial^{2}\delta\dot{\psi}\partial^{2}\delta\dot{\psi}}\partial^{4}
\end{array}\right)\label{hessian}
\end{equation}
is not degenerate, therefore the theory propagates two scalar degrees
of freedom, at least at the linear order in perturbations.

We have shown that the shadowy modes disappear and that there are two physical scalar degrees of freedom. From the point of view of spatial covariance (since we have fixed the gauge $\delta\phi = 0$), one of these two scalar modes corresponds to the breaking of spacetime diffeomorphism to spatial diffeomorphism, while the other simply corresponds to the scalar field that is not chosen to fix the time coordinate.

It is interesting that in the case of $\hat{\mathcal{C}}_{\partial\delta\dot{\psi}\partial\delta\dot{\psi}}=\hat{\mathcal{C}}_{\partial^{2}\delta\dot{\psi}\partial^{2}\delta\dot{\psi}}=0$,
i.e., $f_{7}=f_{8}=f_{9}=0$ or if we focus on the large scales such
that higher order spatial derivatives are sub-dominant, the determinant
of the Hessian matrix reduces to be 
\begin{equation}
\det\bm{H}=4\hat{\mathcal{C}}_{\dot{\zeta}\dot{\zeta}}\hat{\mathcal{C}}_{\delta\dot{\psi}\delta\dot{\psi}}-(\hat{\mathcal{C}}_{\dot{\zeta}\delta\dot{\psi}})^{2}.
\end{equation}
According to the explicit expressions of $\hat{\mathcal{C}}_{\dot{\zeta}\dot{\zeta}}$,
$\hat{\mathcal{C}}_{\delta\dot{\psi}\delta\dot{\psi}}$ and $\hat{\mathcal{C}}_{\dot{\zeta}\delta\dot{\psi}}$
in (\ref{app:Chdzdz}), (\ref{app:Chdzdp}) and (\ref{app:Chdpdp}),
it is interesting that $\hat{\mathcal{C}}_{\dot{\zeta}\delta\dot{\psi}}=\hat{\mathcal{C}}_{\delta\dot{\psi}\delta\dot{\psi}}=0$
if $\mathcal{F}_{1}$ does not depend on $Y$ or $Z$. 
In other words, if the coefficient $\mathcal{F}_{1}$ is only a function of $X$, $\phi$, and $\psi$, the theory would be degenerate, and a single scalar mode (i.e., $\zeta$) would survive. It is also interesting to note that in the case of $\mathcal{F}_{2} = 0$ and/or $\mathcal{F}_{4} = 0$, extra constraints will arise and lead to a further reduction in the number of degrees of freedom. A detailed analysis of the number of scalar degrees of freedom is beyond the scope of our work.

In the following, we briefly discuss the stability of scalar perturbations. To this end, we write the action in momentum space
\begin{eqnarray}
S^{\text{2s}}=\int\text{dt}\text{d}^{3}xL^{\text{2s}} & = & \int\text{d}t\frac{\text{d}^{3}k}{(2\pi)^{3}}\Big[\dot{q}^{A}(t,\boldsymbol{k})\mathcal{K}_{AB}\dot{q}^{B}(t,-\boldsymbol{k})+q^{A}(t,\boldsymbol{k})\mathcal{Q}_{AB}\dot{q}^{B}(t,-\boldsymbol{k})\nonumber \\
 &  & -\frac{\boldsymbol{k}^{2}}{a^{2}}q^{A}(t,\boldsymbol{k})\mathcal{M}_{AB}q^{B}(t,-\boldsymbol{k})-q^{A}(t,\boldsymbol{k})\mathcal{V}_{AB}q^{B}(t,-\boldsymbol{k})\Big],
\end{eqnarray}
where $q^{A}\equiv\left(\begin{array}{cc}
\zeta & \delta\psi\end{array}\right)^{\mathrm{T}}$ with $A=1,2$, and the coefficient matrices are defined as
\begin{align}
\mathcal{K}_{AB} & \coloneqq\begin{pmatrix}\hat{\mathcal{C}}_{\dot{\zeta}\dot{\zeta}} & \frac{1}{2}\hat{\mathcal{C}}_{\dot{\zeta}\delta\dot{\psi}}\\
\frac{1}{2}\hat{\mathcal{C}}_{\dot{\zeta}\delta\dot{\psi}} & \hat{\mathcal{C}}_{\delta\dot{\psi}\delta\dot{\psi}}+\hat{\mathcal{C}}_{\partial\delta\dot{\psi}\partial\delta\dot{\psi}}\boldsymbol{k}^{2}+\hat{\mathcal{C}}_{\partial^{2}\delta\dot{\psi}\partial^{2}\delta\dot{\psi}}\boldsymbol{k}^{4}
\end{pmatrix},\\
\mathcal{Q}_{AB} & \coloneqq\begin{pmatrix}0 & \frac{1}{2}\Big(\hat{\mathcal{C}}_{\zeta\delta\dot{\psi}}-\hat{\mathcal{C}}_{\partial^{2}\zeta\delta\dot{\psi}}\boldsymbol{k}^{2}+\hat{\mathcal{C}}_{\partial^{2}\zeta\partial^{2}\delta\dot{\psi}}\boldsymbol{k}^{4}\Big)\\
\frac{1}{2}\Big(\hat{\mathcal{C}}_{\zeta\delta\dot{\psi}}-\hat{\mathcal{C}}_{\partial^{2}\zeta\delta\dot{\psi}}\boldsymbol{k}^{2}+\hat{\mathcal{C}}_{\partial^{2}\zeta\partial^{2}\delta\dot{\psi}}\boldsymbol{k}^{4}\Big) & 0
\end{pmatrix},\\
\mathcal{M}_{AB} & \coloneqq a^{2}\begin{pmatrix}-\hat{\mathcal{C}}_{\partial\zeta\partial\zeta}-\hat{\mathcal{C}}_{\partial^{2}\zeta\partial^{2}\zeta}\boldsymbol{k}^{2} & \frac{1}{2}\Big(\hat{\mathcal{C}}_{\zeta\partial^{2}\delta\psi}-\hat{\mathcal{C}}_{\partial^{2}\zeta\partial^{2}\delta\psi}\boldsymbol{k}^{2}\Big)\\
\frac{1}{2}\Big(\hat{\mathcal{C}}_{\zeta\partial^{2}\delta\psi}-\hat{\mathcal{C}}_{\partial^{2}\zeta\partial^{2}\delta\psi}\boldsymbol{k}^{2}\Big) & -\hat{\mathcal{C}}_{\partial\delta\psi\partial\delta\psi}-\hat{\mathcal{C}}_{\partial^{2}\delta\psi\partial^{2}\delta\psi}\boldsymbol{k}^{2}
\end{pmatrix},\\
\mathcal{V}_{AB} & \coloneqq\begin{pmatrix}0 & -\frac{1}{2}\hat{\mathcal{C}}_{\zeta\delta\psi}\\
-\frac{1}{2}\hat{\mathcal{C}}_{\zeta\delta\psi} & -\hat{\mathcal{C}}_{\delta\psi\delta\psi}
\end{pmatrix}.
\end{align}
The stability of scalar perturbations requires that the matrices $\mathcal{K}_{AB}$
and $\mathcal{M}_{AB}$ must be positive definite. The two eigenvalues
of $\mathcal{K}_{AB}$ are 
\begin{eqnarray}
\lambda_{\pm}^{\mathcal{K}} & = & \frac{1}{2}\Big\{\hat{\mathcal{C}}_{\dot{\zeta}\dot{\zeta}}+\bigl(\hat{\mathcal{C}}_{\delta\dot{\psi}\delta\dot{\psi}}+\boldsymbol{k}^{2}\hat{\mathcal{C}}_{\partial\delta\dot{\psi}\partial\delta\dot{\psi}}+\boldsymbol{k}^{4}\hat{\mathcal{C}}_{\partial^{2}\delta\dot{\psi}\partial^{2}\delta\dot{\psi}}\bigr)\pm\Big[\hat{\mathcal{C}}_{\dot{\zeta}\dot{\zeta}}^{2}+\hat{\mathcal{C}}_{\dot{\zeta}\delta\dot{\psi}}^{2}\nonumber \\
 &  & \quad-2\hat{\mathcal{C}}_{\dot{\zeta}\dot{\zeta}}\bigl(\hat{\mathcal{C}}_{\delta\dot{\psi}\delta\dot{\psi}}+\boldsymbol{k}^{2}\hat{\mathcal{C}}_{\partial\delta\dot{\psi}\partial\delta\dot{\psi}}+\boldsymbol{k}^{4}\hat{\mathcal{C}}_{\partial^{2}\delta\dot{\psi}\partial^{2}\delta\dot{\psi}}\bigr)\nonumber \\
 &  & \quad+\Big(\hat{\mathcal{C}}_{\delta\dot{\psi}\delta\dot{\psi}}+\boldsymbol{k}^{2}\hat{\mathcal{C}}_{\partial\delta\dot{\psi}\partial\delta\dot{\psi}}+\boldsymbol{k}^{4}\hat{\mathcal{C}}_{\partial^{2}\delta\dot{\psi}\partial^{2}\delta\dot{\psi}}\Big)^{2}\Big]^{1/2}\Big\},
\end{eqnarray}
which implies that we must require $\lambda_{\pm}^{\mathcal{K}}>0$
to evade the ghost instability. On the other hand, the two eigenvalues
of $\mathcal{M}_{AB}$ are 
\begin{eqnarray}
\lambda_{\pm}^{\mathcal{M}} & = & \frac{a^{2}}{2}\Big\{-\hat{\mathcal{C}}_{\partial\zeta\partial\zeta}-\boldsymbol{k}^{2}\hat{\mathcal{C}}_{\partial^{2}\zeta\partial^{2}\zeta}-\hat{\mathcal{C}}_{\partial\delta\psi\partial\delta\psi}-\boldsymbol{k}^{2}\hat{\mathcal{C}}_{\partial^{2}\delta\psi\partial^{2}\delta\psi}\pm\Big[(\hat{\mathcal{C}}_{\partial\zeta\partial\zeta}+\boldsymbol{k}^{2}\hat{\mathcal{C}}_{\partial^{2}\zeta\partial^{2}\zeta})^{2}\nonumber \\
 &  & \quad+(\hat{\mathcal{C}}_{\zeta\partial^{2}\delta\psi}-\boldsymbol{k}^{2}\hat{\mathcal{C}}_{\partial^{2}\zeta\partial^{2}\delta\psi})^{2}-2(\hat{\mathcal{C}}_{\partial\zeta\partial\zeta}+\boldsymbol{k}^{2}\hat{\mathcal{C}}_{\partial^{2}\zeta\partial^{2}\zeta})(\hat{\mathcal{C}}_{\partial\delta\psi\partial\delta\psi}+\boldsymbol{k}^{2}\hat{\mathcal{C}}_{\partial^{2}\delta\psi\partial^{2}\delta\psi})\nonumber \\
 &  & \quad+(\hat{\mathcal{C}}_{\partial\delta\psi\partial\delta\psi}+\boldsymbol{k}^{2}\hat{\mathcal{C}}_{\partial^{2}\delta\psi\partial^{2}\delta\psi})^{2}\Big]^{1/2}\Big\}.
\end{eqnarray}
Thus we must require $\lambda_{\pm}^{\mathcal{M}}>0$ to evade the
gradient instability.

\subsection{The linear tensor and vector perturbations}

In the above, we have seen that our theory indeed propagates at most two scalar modes at the linear order in perturbations. For completeness, we will briefly discuss the tensor and vector perturbations in the following.

The quadratic action for the tensor modes is given by
\begin{equation}
S^{\text{2t}}=\int\text{d}t\text{\ensuremath{\frac{\text{d}^{3}k}{(2\pi)^{3}}}}\frac{a^{3}}{2}\mathcal{G}\bigg(\partial_{t}\gamma_{ij}(t,\boldsymbol{k})\partial_{t}\gamma^{ij}(t,-\boldsymbol{k})-\frac{\boldsymbol{k}^{2}}{a^{2}}\frac{\mathcal{W}}{\mathcal{G}}\gamma_{ij}(t,\boldsymbol{k})\gamma^{ij}(t,-\boldsymbol{k})\bigg),
\end{equation}
with 
\begin{eqnarray}
\mathcal{G} & = & \frac{1}{2}(f_{1}+f_{3}+f_{5}),\\
\mathcal{W} & = & \frac{1}{2}(f_{10}+f_{11}).
\end{eqnarray}
We should require $\mathcal{G} > 0$ in order to evade the ghost modes and $\mathcal{W} > 0$ for gradient stability. The propagation speed of gravitational waves is thus determined by $c_{\text{GW}}^{2} = \mathcal{W}/\mathcal{G}$.

Straightforward calculations yield the quadratic Lagrangian for the vector modes
\begin{equation}
L^{\text{2v}}=\mathcal{G}a^{3}\partial_{i}\dot{F}_{j}\partial^{i}\dot{F}^{j}-\mathcal{G}a^{2}\partial_{i}\dot{F}_{j}\partial^{i}S^{j}+\mathcal{G}a\partial_{i}S_{j}\partial^{i}S^{j}.
\end{equation}
The gauge invariant variables for the vector modes, i.e., $a\dot{F}_{i}-S_{i}$,
satisfy the constraint equation
\begin{equation}
\partial^{2}(a\dot{F}_{i}-S_{i})=0,
\end{equation}
which is exactly the same as that of general relativity. As a result,
there is no vector mode propagating in our theory. 

To conclude, there are generally (at most) two scalar and two tensor
degrees of freedom propagating at the linear order in perturbations
in our theory. 

\section{Conclusion \label{sec:Conclusion}}

Recently, many attempts have been made to pursue general multi-scalar-tensor theories without ghost instabilities or unwanted degrees of freedom. It has proven difficult to generalize the fine-tuning approach from the case of a single scalar field to multiple scalar fields \citep{Ohashi:2015fma,Akama:2017jsa} due to the complexities of couplings among different fields. Until recently, Horndeski attempted to construct the most general bi-scalar-tensor theory with second-order equations of motion in four dimensions \citep{Horndeski:2024hee}. However, the development of single-field scalar-tensor theories inspires researchers to search for theories beyond Galileon, particularly from the perspective of degeneracy or background-dependent models.

The spatially covariant gravity theory, which can be viewed as the gauge-fixed version of the generally covariant scalar-tensor theory, provides an alternative approach to constructing healthy scalar-tensor theories. The basic idea is to use geometric quantities of hypersurfaces specified by the scalar field to develop the theory. In this work, we generalize this idea to the case of two scalar fields. In Sec. \ref{sec:geo}, we describe the geometric framework of our scenario, in which each scalar field specifies a foliation of spacelike hypersurfaces, accompanied by its adapted hypersurface geometric quantities. Thus, we have two independent sets of geometric quantities as our building blocks, which are summarized in Table \ref{tab:blocks}. In Sec. \ref{sec:action}, we construct the action based on these hypersurface geometric quantities. The most general action is given in (\ref{S_gen}). To evade ghost-like or unwanted degrees of freedom, we made some additional assumptions about the structure of the action. In particular, we require that there should be no ``explicit'' mixing derivative terms and no higher-order temporal/Lie derivative terms. The resulting action thus takes the form of (\ref{S_fin}). In Sec. \ref{sec:mono}, we exhaustively list all possible monomials without parity violation, i.e., scalar contractions, in the form of (\ref{S_fin}) up to $d=3$, where $d$ is the total number of derivatives in each monomial. These monomials are listed in Table \ref{tab:Mono1} for $d=1$, in Table \ref{tab:mono2} for $d=2$, and in Tables \ref{tab:mono3_1}-\ref{tab:factmono3_3} for $d=3$. Notably, nontrivial monomials such as $a_{a}v^{a}$, $v^{a}\mathrm{D}_{a}K$, and $v^{a}v^{b}\mathrm{D}_{a}a_{b}$ arise, which do not exist or can be reduced in the case of a single scalar field. Having constructed the Lagrangian in terms of hypersurface geometric quantities, it is crucial to derive the corresponding expressions of the bi-scalar-tensor theory, as discussed in \citep{Gao:2020qxy,Gao:2020yzr,Hu:2021bbo}, which we perform in Sec. \ref{sec:corr}. In Sec. \ref{sec:pert}, by employing a perturbative analysis around a cosmological background and using a simple model (\ref{Lagd2}) as an illustration, we explicitly demonstrate that the bi-scalar-tensor theory constructed using our method indeed propagates two tensor and two scalar degrees of freedom at the linear order in perturbations, thereby ensuring it is free of ghost or unwanted degrees of freedom.

Although our construction is justified through a cosmological perturbation analysis up to the linear order, it is important to prove that it is free of extra degrees of freedom at the nonlinear level, for example, through a rigorous Hamiltonian analysis. There are also several possible extensions of the results presented in this work. First, we assume that Lie derivatives enter the Lagrangian only through the extrinsic curvatures; however, one might extend the action (\ref{S_fin}) by including the first-order Lie derivatives of the normalization factors, i.e., $\pounds_{\bm{u}}N$ and $\pounds_{\bm{v}}M$, while maintaining the number of degrees of freedom, as analyzed in \citep{Gao:2018znj,Lin:2020nro}. Second, we have considered only parity-preserving monomials; one may also investigate parity-violating monomials and their scalar-tensor theory correspondence by generalizing the results in \citep{Hu:2024hzo}. Moreover, it was shown in \citep{Gao:2019lpz} that a more general spatially covariant gravity could be equivalent to a simpler one through some invertible field transformations. Thus, it is essential to check whether the bi-scalar-tensor theory obtained through our construction can be related to the known ones through any field transformations. We will leave these issues for future work.

\begin{acknowledgments}
This work was partly supported by the National Natural Science Foundation
of China (NSFC) under Grants No. 12475068 and No. 11975020 (X. G.)
and No. 12347137 (Y. M. H.), and the China Postdoctoral Science Foundation
under Grant No. 2024M753076 (Y. M. H.).
\end{acknowledgments}

\appendix

\section{Factorizable monomials of $d=3$ \label{app:mono3}}

In this appendix, we list the factorizable monomials of $d=3$ in
Tables \ref{tab:factmono3_1}, \ref{tab:factmono3_2}, \ref{tab:factmono3_3}.

\begin{table}[H]
\begin{centering}
\begin{tabular}{|c|>{\centering}p{6cm}|>{\centering}p{5cm}|}
\hline 
B.b. & Monomials: $[2+1]$ & Monomials: $[1+1+1]$\tabularnewline
\hline 
$KKK$ & $[K][KK]$, $[K][KvKv]$, $[Kvv][KK]$, $[Kvv][KvKv]$ & $[K]^{3}$, $[K]^{2}[Kvv]$, $[K][Kvv]^{2}$, $[Kvv]^{3}$\tabularnewline
\hline 
$KKa$ & $[K][Kva]$, $[Kvv][Kva]$, $[av][KK]$, $[av][KvKv]$ & $[K]^{2}[av]$, $[K][Kvv][av]$, $[Kvv]^{2}[av]$\tabularnewline
\hline 
$Kaa$ & $[K][aa]$, $[Kvv][aa]$, $[av][Kva]$ & $[K][av]^{2}$, $[Kvv][av]^{2}$\tabularnewline
\hline 
$K^{3}\!R$ & $[K][^{3}\!R]$, $[K][^{3}\!Rvv]$, $[Kvv][^{3}\!R]$, $[Kvv][^{3}\!Rvv]$ & -\tabularnewline
\hline 
$K\text{D}K$ & $[K][v\text{D}K]$, $[K][\text{D}Kv]$, $[K][v\text{D}Kvv]$, $[Kvv][v\text{D}K]$,
$[Kvv][\text{D}Kv]$, $[Kvv][v\text{D}Kvv]$ & -\tabularnewline
\hline 
$K\text{D}a$ & $[K][\text{D}a]$, $[K][v\text{D}av]$, $[Kvv][\text{D}a]$, $[Kvv][v\text{D}av]$ & -\tabularnewline
\hline 
$aaa$ & $[av][aa]$ & $[av]^{3}$\tabularnewline
\hline 
$a^{3}\!R$ & $[av][^{3}\!R]$, $[av][^{3}\!Rvv]$ & -\tabularnewline
\hline 
$a\text{D}a$ & $[av][\text{D}a]$, $[av][v\text{D}av]$ & -\tabularnewline
\hline 
\end{tabular}
\par\end{centering}
\caption{Factorizable monomials of $d=3$: part I.}

\label{tab:factmono3_1}
\end{table}

\begin{table}[H]
\begin{centering}
\begin{tabular}{|c|>{\centering}p{6cm}|>{\centering}p{5cm}|}
\hline 
B.b. & Monomials: $[2+1]$ & Monomials: $[1+1+1]$\tabularnewline
\hline 
$LLL$ & $[L][LL]$, $[L][LuLu]$, $[Luu][LL]$, $[Luu][LuLu]$ & $[L]^{3}$, $[L]^{2}[Luu]$, $[L][Luu]^{2}$, $[Luu]^{3}$\tabularnewline
\hline 
$LLb$ & $[L][Lub]$, $[Lh][Lub]$, $[bu][LL]$, $[bu][LuLu]$ & $[L]^{2}[bu]$, $[L][Lu][bu]$, $[Lu]^{2}[bu]$\tabularnewline
\hline 
$Lbb$ & $[L][bb]$, $[Luu][bb]$, $[Lua][bu]$ & $[L][bu]^{2}$, $[Luu][bu]^{2}$\tabularnewline
\hline 
$L^{3}\!\tilde{R}$ & $[L][^{3}\!\tilde{R}]$, $[L][^{3}\!\tilde{R}uu]$, $[Luu][^{3}\!\tilde{R}]$,
$[Luu][^{3}\!\tilde{R}uu]$ & -\tabularnewline
\hline 
$L\tilde{\text{D}}L$ & $[L][u\tilde{\text{D}}L]$, $[L][\tilde{\text{D}}Lu]$, $[L][u\tilde{\text{D}}Luu]$,
$[Luu][u\tilde{\text{D}}L]$, $[Luu][\tilde{\text{D}}Lu]$, $[Luu][u\tilde{\text{D}}Luu]$ & -\tabularnewline
\hline 
$L\tilde{\text{D}}b$ & $[L][\tilde{\text{D}}b]$, $[L][u\tilde{\text{D}}bu]$, $[Luu][\tilde{\text{D}}b]$,
$[Luu][u\tilde{\text{D}}bu]$ & -\tabularnewline
\hline 
$bbb$ & $[bu][bb]$ & $[bu]^{3}$\tabularnewline
\hline 
$b^{3}\!\tilde{R}$ & $[bu][^{3}\!\tilde{R}]$, $[bu][^{3}\!\tilde{R}uu]$ & -\tabularnewline
\hline 
$b\tilde{\text{D}}b$ & $[bu][\tilde{\text{D}}b]$, $[bu][u\tilde{\text{D}}bu]$ & -\tabularnewline
\hline 
\end{tabular}
\par\end{centering}
\caption{Factorizable monomials of $d=3$: part II.}

\label{tab:factmono3_2}
\end{table}

\begin{table}[H]
\begin{centering}
\begin{tabular}{|c|>{\centering}p{6cm}|>{\centering}p{5cm}|}
\hline 
B.b. & Monomials: $[2+1]$ & Monomials: $[1+1+1]$\tabularnewline
\hline 
$KKL$ & $[K][KL]$, $[K][KvLu]$, $[Kvv][KL]$, $[Kvv][KvLu]$, $[KK][L]$,
$[KvKv][L]$, $[KK][Luu]$, $[KvKv][Luu]$ & $[K]^{2}[L]$, $[K]^{2}[Luu]$, $[K][Kvv][L]$, $[K][Kvv][Luu]$,
$[Kvv]^{2}[L]$, $[Kvv]^{2}[Luu]$\tabularnewline
\hline 
$KKb$ & $[K][Kvb]$, $[Kvv][Kvb]$, $[bu][KK]$, $[bu][KvKv]$ & $[K]^{2}[bu]$, $[K][Kvv][bu]$, $[Kvv]^{2}[bu]$\tabularnewline
\hline 
$KLL$ & $[K][LL]$, $[K][LhL]$, $[Kvv][LL]$, $[Kvv][LhL]$, $[L][KL]$,
$[L][KvLu]$, $[Luu][KL]$, $[Luu][KvLu]$ & $[L]^{2}[K]$, $[L]^{2}[Kvv]$, $[L][Lh][K]$, $[L][Lh][Kvv]$, $[Lh]^{2}[K]$,
$[Lh]^{2}[Kvv]$\tabularnewline
\hline 
$KLa$ & $[K][Lua]$, $[Kvv][Lua]$, $[L][Kva]$, $[Luu][Kva]$, $[av][KL]$,
$[av][KvLu]$ & $[K][L][av]$, $[K][Luu][av]$, $[Kvv][L][av]$, $[Kvv][Luu][av]$\tabularnewline
\hline 
$KLb$ & $[K][Lub]$, $[Kvv][Lub]$, $[L][Kvb]$, $[Luu][Kvb]$, $[bu][KL]$,
$[bu][KvLu]$ & $[K][L][bu]$, $[K][Luu][bu]$, $[Kvv][L][bu]$, $[Kvv][Luu][bu]$\tabularnewline
\hline 
$Kab$ & $[K][ab]$, $[Kvv][ab]$, $[av][Kvb]$, $[bu][Kva]$ & $[K][av][bu]$, $[Kvv][av][bu]$\tabularnewline
\hline 
$Kbb$ & $[K][bb]$, $[Kvv][bb]$, $[bu][Kvb]$ & $[K][bu]^{2}$, $[Kvv][bu]^{2}$\tabularnewline
\hline 
$K^{3}\tilde{R}$ & $[K][^{3}\tilde{R}]$, $[K][^{3}\tilde{R}uu]$, $[Kvv][^{3}\tilde{R}]$,
$[Kvv][^{3}\tilde{R}uu]$ & -\tabularnewline
\hline 
$K\tilde{\text{D}}L$ & $[K][u\tilde{\text{D}}L]$, $[K][\tilde{\text{D}}Lu]$, $[K][u\tilde{\text{D}}Luu]$,
$[Kvv][u\tilde{\text{D}}L]$, $[Kvv][\tilde{\text{D}}Lu]$, $[Kvv][u\tilde{\text{D}}Luu]$ & -\tabularnewline
\hline 
$K\tilde{\text{D}}b$ & $[K][\tilde{\text{D}}b]$, $[K][u\tilde{\text{D}}bu]$, $[Kvv][\tilde{\text{D}}b]$,
$[Kvv][u\tilde{\text{D}}bu]$ & -\tabularnewline
\hline 
$LLa$ & $[L][Lua]$, $[Lh][Lua]$, $[av][LL]$, $[av][LuLu]$ & $[L]^{2}[av]$, $[L][Lu][av]$, $[Lu]^{2}[av]$\tabularnewline
\hline 
$Laa$ & $[L][aa]$, $[Luu][aa]$, $[Lua][av]$ & $[L][av]^{2}$, $[Luu][av]^{2}$\tabularnewline
\hline 
$Lab$ & $[L][ab]$, $[Luu][ab]$, $[av][Lub]$, $[bu][Lua]$ & $[L][av][bu]$, $[Luu][av][bu]$\tabularnewline
\hline 
$L^{3}R$ & $[L][^{3}R]$, $[L][^{3}Rvv]$, $[Luu][^{3}R]$, $[Luu][^{3}Rvv]$ & -\tabularnewline
\hline 
$L\text{D}a$ & $[L][\text{D}a]$, $[L][v\text{D}av]$, $[Luu][\text{D}a]$, $[Luu][v\text{D}av]$ & -\tabularnewline
\hline 
$aab$ & $[av][ab]$, $[bu][aa]$ & $[av]^{2}[bu]$\tabularnewline
\hline 
$abb$ & $[av][bb]$, $[bu][ab]$ & $[av][bu]^{2}$\tabularnewline
\hline 
$a^{3}\tilde{R}$ & $[av][^{3}\tilde{R}]$, $[av][^{3}\tilde{R}uu]$ & -\tabularnewline
\hline 
$a\tilde{\text{D}}b$ & $[av][\tilde{\text{D}}b]$, $[av][u\tilde{\text{D}}bu]$ & -\tabularnewline
\hline 
$b^{3}R$ & $[bu][^{3}R]$, $[bu][^{3}Rvv]$ & -\tabularnewline
\hline 
\end{tabular}
\par\end{centering}
\caption{Factorizable monomials of $d=3$: part III.}

\label{tab:factmono3_3}
\end{table}

\section{Coefficients in the quadratic Lagrangian for the scalar perturbations
\label{app:coeff}}

The quadratic Lagrangian for the scalar perturbations $A$, $B$,
$\zeta$ and $\delta\psi$ takes the form
\begin{eqnarray}
L^{\text{2s}} & \simeq & \mathcal{C}_{\dot{\zeta}\dot{\zeta}}\dot{\zeta}^{2}+\mathcal{C}_{\dot{\zeta}\delta\dot{\psi}}\dot{\zeta}\delta\dot{\psi}+\mathcal{C}_{\dot{\zeta}\partial^{2}B}\dot{\zeta}\partial^{2}B+\mathcal{C}_{\delta\dot{\psi}\delta\dot{\psi}}\delta\dot{\psi}^{2}+\mathcal{C}_{\partial\delta\dot{\psi}\partial\delta\dot{\psi}}\partial_{i}\delta\dot{\psi}\partial^{i}\delta\dot{\psi}\nonumber \\
 &  & +\mathcal{C}_{\delta\dot{\psi}A}\delta\dot{\psi}A+\mathcal{C}_{\partial\delta\dot{\psi}\partial A}\partial^{i}\delta\dot{\psi}\partial_{i}A+\mathcal{C}_{\delta\dot{\psi}\partial^{2}B}\delta\dot{\psi}\partial^{2}B+\mathcal{C}_{AA}A^{2}\nonumber \\
 &  & +\mathcal{C}_{\partial A\partial A}\partial^{i}A\partial_{i}A+\mathcal{C}_{\partial^{2}B\partial^{2}B}\partial^{2}B\partial^{2}B\nonumber \\
 &  & +\mathcal{C}_{\partial\zeta\partial\zeta}\partial^{i}\zeta\partial_{i}\zeta+\mathcal{C}_{\zeta\delta\dot{\psi}}\zeta\delta\dot{\psi}+\mathcal{C}_{\partial^{2}\zeta\delta\dot{\psi}}\partial^{2}\zeta\delta\dot{\psi}+\mathcal{C}_{\zeta\delta\psi}\zeta\delta\psi+\mathcal{C}_{\zeta\partial^{2}\delta\psi}\zeta\partial^{2}\delta\psi\nonumber \\
 &  & +\mathcal{C}_{\partial^{2}\zeta A}\partial^{2}\zeta A+\mathcal{C}_{\delta\psi\delta\psi}\delta\psi^{2}+\mathcal{C}_{\partial\delta\psi\partial\delta\psi}\partial^{i}\delta\psi\partial_{i}\delta\psi+\mathcal{C}_{\partial^{2}\delta\psi\partial^{2}\delta\psi}\partial^{2}\delta\psi\partial^{2}\delta\psi\nonumber \\
 &  & +\mathcal{C}_{\delta\psi A}\delta\psi A+\mathcal{C}_{\partial^{2}\delta\psi A}\partial^{2}\delta\psi A+\mathcal{C}_{\delta\psi\partial^{2}B}\delta\psi\partial^{2}B+\mathcal{C}_{\partial^{2}\delta\psi\partial^{2}B}\partial^{2}\delta\psi\partial^{2}B.\label{eq:order-2 scalar action-1}
\end{eqnarray}
where the explicit coefficients for are listed as follows. The coefficients relevant to the kinetic terms of $\zeta$ and $\delta\psi$
are
\begin{eqnarray}
\mathcal{C}_{\dot{\zeta}\dot{\zeta}} & = & 3a^{3}\mathcal{F}_{1},\\
\mathcal{C}_{\partial\delta\dot{\psi}\partial\delta\dot{\psi}} & = & af_{9}/\dot{\bar{\psi}}^{2},\\
\mathcal{C}_{\dot{\zeta}\delta\dot{\psi}} & = & 3a^{3}H(\mathcal{F}_{1Z}\dot{\bar{\phi}}+2\mathcal{F}_{1Y}\dot{\bar{\psi}}),\\
\mathcal{C}_{\dot{\zeta}\partial^{2}B} & = & -2a^{2}\mathcal{F}_{1},\\
\mathcal{C}_{\delta\dot{\psi}\delta\dot{\psi}} & = & \frac{3}{8}a^{3}H^{2}\Big[\mathcal{F}_{1ZZ}\dot{\bar{\phi}}^{2}+4\Big(\mathcal{F}_{1Y}+\mathcal{F}_{1YY}\dot{\bar{\psi}}^{2}+\mathcal{F}_{1YZ}\dot{\bar{\phi}}\dot{\bar{\psi}}\Big)\Big],\\
\mathcal{C}_{\delta\dot{\psi}A} & = & -\frac{3}{2}a^{3}H^{2}\Big[\mathcal{F}_{1XZ}\dot{\bar{\phi}}^{3}+2\mathcal{F}_{1YY}\dot{\bar{\psi}}^{3}+3\mathcal{F}_{1Z}\dot{\bar{\phi}}\nonumber \\
 &  & +3\mathcal{F}_{1YZ}\dot{\bar{\psi}}^{2}\dot{\bar{\phi}}+6\mathcal{F}_{1Y}\dot{\bar{\psi}}+(\mathcal{F}_{1ZZ}+2\mathcal{F}_{1XY})\dot{\bar{\phi}}^{2}\dot{\bar{\psi}}\Big],\\
\mathcal{C}_{\partial\delta\dot{\psi}\partial A} & = & -a\left(f_{8}+2f_{9}\right)/\dot{\bar{\psi}},\\
\mathcal{C}_{\delta\dot{\psi}\partial^{2}B} & = & -a^2H(\mathcal{F}_{1Z}\dot{\bar{\phi}}+2\mathcal{F}_{1Y}\dot{\bar{\psi}}),
\end{eqnarray}
as well as
\begin{eqnarray}
\mathcal{C}_{AA} & = & \frac{3}{2}a^{3}H^{2}\Big[\mathcal{F}_{1}+4\mathcal{F}_{1X}\dot{\bar{\phi}}^{2}+4\mathcal{F}_{1Y}\dot{\bar{\psi}}^{2}+\mathcal{F}_{1XX}\dot{\bar{\phi}}^{4}+\mathcal{F}_{1YY}\dot{\bar{\psi}}^{4}\nonumber \\
 &  & \quad+(\mathcal{F}_{1ZZ}+2\mathcal{F}_{1XY})\dot{\bar{\psi}}^{4}+2\mathcal{F}_{1YZ}\dot{\bar{\psi}}^{3}\dot{\bar{\phi}}\nonumber \\
 &  & \quad+2\mathcal{F}_{1XZ}\dot{\bar{\phi}}^{3}\dot{\bar{\psi}}+4\mathcal{F}_{1Z}\dot{\bar{\phi}}\dot{\bar{\psi}}\Big],\\
\mathcal{C}_{\partial A\partial A} & = & a(f_{7}+f_{8}+f_{9}),\\
\mathcal{C}_{\partial^{2}B\partial^{2}B} & = & a\mathcal{F}_{2}.
\end{eqnarray}
Other coefficients (irrelevant to the kinetic terms) are
\begin{eqnarray}
\mathcal{C}_{\partial\zeta\partial\zeta} & = & 2a(f_{10}+f_{11}),\\
\mathcal{C}_{\zeta\delta\dot{\psi}} & = & -6a^{3}H\mathcal{F}_{1\psi}+\frac{9}{2}a^{3}H^{2}(\mathcal{F}_{1Z}\dot{\bar{\phi}}+2\mathcal{F}_{1Y}\dot{\bar{\psi}}),\\
\mathcal{C}_{\partial^{2}\zeta\delta\dot{\psi}} & = & \frac{a(4f_{11}+\mathcal{F}_{3})}{\dot{\bar{\psi}}}-2a(f_{10Z}+f_{11Z})\dot{\bar{\phi}}-4a(f_{10Y}+f_{11Y})\dot{\bar{\psi}},\\
\mathcal{C}_{\zeta\delta\psi} & = & -3a^{3}H^{2}\mathcal{F}_{1\psi}-6a^{2}(aH\dot{\mathcal{F}}_{1\psi}+\ddot{a}\mathcal{F}_{1\psi}),\\
\mathcal{C}_{\zeta\partial^{2}\delta\psi} & = & -4a(f_{10\psi}+f_{11\psi})-\frac{4a}{\dot{\bar{\psi}}^{2}}(f_{11}+\mathcal{F}_{3})\ddot{\bar{\psi}}\nonumber \\
 &  & +\frac{1}{\dot{\bar{\psi}}}\Big[a(4\dot{f}_{11}+\dot{\mathcal{F}}_{3})+aH(4f_{11}+\mathcal{F}_{3})\Big],\\
\mathcal{C}_{\partial^{2}\zeta A} & = & 4a\Big[-(f_{10}+f_{11})+(f_{10X}+f_{11X})\dot{\bar{\phi}}^{2}\nonumber \\
 &  & \quad+(f_{10Y}+f_{11Y})\dot{\bar{\psi}}^{2}+(f_{10Z}+f_{11Z})\dot{\bar{\phi}}\dot{\bar{\psi}}\Big],\\
\mathcal{C}_{\delta\psi\delta\psi} & = & -\frac{3}{4}\Big[a^{3}H^{3}(\mathcal{F}_{1\psi Z}\dot{\bar{\phi}}+2\mathcal{F}_{1\psi Y}\dot{\bar{\psi}})+2a^{2}H\ddot{a}(\mathcal{F}_{1\psi Z}\dot{\bar{\phi}}+2\mathcal{F}_{1\psi Y}\dot{\bar{\psi}})\nonumber \\
 &  & \quad+a^{3}H^{2}(-2\mathcal{F}_{1\psi\psi}+\dot{\mathcal{F}}_{1}\dot{\bar{\phi}}+2\dot{\mathcal{F}}_{1\psi Y}\dot{\bar{\psi}}+\mathcal{F}_{1\psi Z}\ddot{\bar{\phi}}+2\mathcal{F}_{1\psi Y}\ddot{\bar{\psi}})\Big],\\
\mathcal{C}_{\partial\delta\psi\partial\delta\psi} & = & \frac{aH^{2}}{2\dot{\bar{\psi}}^{2}}\Big(4f_{11}+f_{3}+\mathcal{F}_{3}-3\mathcal{F}_{1Y}\dot{\bar{\psi}}^{2}\Big)\nonumber \\
 &  & -2af_{9}\frac{\ddot{\bar{\psi}}^{2}}{\dot{\bar{\psi}}^{4}}+\frac{1}{4}\left[8aH\dot{f}_{11Y}-2aH\dot{\mathcal{F}}_{3Y}-2\ddot{a}(-4f_{11Y}+\mathcal{F}_{3})\right]\nonumber \\
 &  & +\frac{1}{4\dot{\bar{\psi}}}\Big[\ddot{a}(-4f_{11Z}+\mathcal{F}_{3Z})\dot{\bar{\phi}}+aH(16f_{11\psi}-4\mathcal{F}_{3}-4\dot{f}_{11Z}\dot{\bar{\phi}}\nonumber \\
 &  & \quad+\dot{\mathcal{F}}_{3}\dot{\bar{\phi}}-4f_{11Z}\ddot{\bar{\phi}}+\mathcal{F}_{3Z}\ddot{\bar{\phi}})\Big]\nonumber \\
 &  & -\frac{1}{4\dot{\bar{\psi}}^{2}}\Big(8aH\dot{f}_{11}-2aH\dot{\mathcal{F}}_{3}-8\ddot{a}f_{11}-2\ddot{a}\mathcal{F}_{3}-aH\left(4f_{11Z}-\mathcal{F}_{3Z}\right)\dot{\bar{\phi}}\ddot{\bar{\psi}}\Big)\nonumber \\
 &  & -\frac{1}{\dot{\bar{\psi}}^{3}}\Big(a\dot{f}_{9}\ddot{\bar{\psi}}+aHf_{9}\ddot{\bar{\psi}}+af_{9}\dddot{\bar{\psi}}\Big),\\
\mathcal{C}_{\partial^{2}\delta\psi\partial^{2}\delta\psi} & = & \frac{f_{5}+f_{6}}{a\dot{\bar{\psi}}^{2}},\\
\mathcal{C}_{\delta\psi A} & = & -3a^{3}H^{2}\Big(\mathcal{F}_{1\psi}+\mathcal{F}_{1\psi X}\dot{\bar{\phi}}^{2}+\mathcal{F}_{1\psi Z}\dot{\bar{\phi}}\dot{\bar{\psi}}+\mathcal{F}_{1\psi Y}\dot{\bar{\psi}}^{2}\Big),\\
\mathcal{C}_{\partial^{2}\delta\psi A} & = & -\frac{a}{\dot{\bar{\psi}}^{2}}(f_{8}+2f_{9})\ddot{\bar{\psi}}+\frac{aH}{\dot{\bar{\psi}}}\Big[4f_{11}+\mathcal{F}_{3}+(-4f_{11X}+\mathcal{F}_{3X})\dot{\bar{\phi}}^{2}\Big]\nonumber \\
 &  & +\Big[(-4f_{11Z}+\mathcal{F}_{3Z})\dot{\bar{\phi}}+(-4f_{11Y}+\mathcal{F}_{3Y})\dot{\bar{\psi}}\bigr)\Big],\\
\mathcal{C}_{\delta\psi\partial^{2}B} & = & -2a^{2}H\mathcal{F}_{1\psi}+\frac{3}{2}a^{2}H^{2}(\mathcal{F}_{1Z}\dot{\bar{\phi}}+2\mathcal{F}_{1Y}\dot{\bar{\psi}}),\\
\mathcal{C}_{\partial^{2}\delta\psi\partial^{2}B} & = & \frac{1}{\dot{\bar{\psi}}}(f_{3}+f_{4}+2f_{5}+2f_{6}),
\end{eqnarray}
with 
\begin{eqnarray}
\mathcal{F}_{1} & \coloneqq & f_{1}+3f_{2}+f_{3}+3f_{4}+f_{5}+3f_{6},\\
\mathcal{F}_{2} & \coloneqq & f_{1}+f_{2}+f_{3}+f_{4}+f_{5}+f_{6},\\
\mathcal{F}_{3} & \coloneqq & f_{3}+3f_{4}+2f_{5}+6f_{6}.
\end{eqnarray}
After plugging the solutions for the auxiliary variables $A$ and
$B$ into the quadratic Lagrangian for the perturbations, and assuming
that $f_{7}+f_{8}+f_{9}=0$, the effective Lagrangian for $\zeta$,
$\delta\psi$ takes the form
\begin{eqnarray}
L^{\text{2s}} & = & \hat{\mathcal{C}}_{\dot{\zeta}\dot{\zeta}}\dot{\zeta}^{2}+\hat{\mathcal{C}}_{\dot{\zeta}\delta\dot{\psi}}\dot{\zeta}\delta\dot{\psi}+\hat{\mathcal{C}}_{\delta\dot{\psi}\delta\dot{\psi}}\delta\dot{\psi}^{2}+\hat{\mathcal{C}}_{\partial\delta\dot{\psi}\partial\delta\dot{\psi}}\partial_{i}\delta\dot{\psi}\partial^{i}\delta\dot{\psi}+\hat{\mathcal{C}}_{\partial^{2}\delta\dot{\psi}\partial^{2}\delta\dot{\psi}}(\partial^{2}\delta\dot{\psi})^{2}\nonumber \\
 &  & +\hat{\mathcal{C}}_{\zeta\delta\dot{\psi}}\zeta\delta\dot{\psi}+\hat{\mathcal{C}}_{\partial^{2}\zeta\delta\dot{\psi}}\partial^{2}\zeta\delta\dot{\psi}+\hat{\mathcal{C}}_{\partial^{2}\zeta\partial^{2}\delta\dot{\psi}}\partial^{2}\zeta\partial^{2}\delta\dot{\psi}\nonumber \\
 &  & +\hat{\mathcal{C}}_{\partial\zeta\partial\zeta}\partial^{i}\zeta\partial_{i}\zeta+\hat{\mathcal{C}}_{\partial^{2}\zeta\partial^{2}\zeta}(\partial^{2}\zeta)^{2}+\hat{\mathcal{C}}_{\zeta\delta\psi}\zeta\delta\psi+\hat{\mathcal{C}}_{\zeta\partial^{2}\delta\psi}\zeta\partial^{2}\delta\psi+\hat{\mathcal{C}}_{\partial^{2}\zeta\partial^{2}\delta\psi}\partial^{2}\zeta\partial^{2}\delta\psi\nonumber \\
 &  & +\hat{\mathcal{C}}_{\delta\psi\delta\psi}(\delta\psi)^{2}+\hat{\mathcal{C}}_{\partial\delta\psi\partial\delta\psi}\partial^{i}\delta\psi\partial_{i}\delta\psi+\hat{\mathcal{C}}_{\partial^{2}\delta\psi\partial^{2}\delta\psi}(\partial^{2}\delta\psi)^{2},
\end{eqnarray}
where the coefficients are
\begin{eqnarray}
\hat{\mathcal{C}}_{\dot{\zeta}\dot{\zeta}} & = & -\frac{a^{3}}{\mathcal{F}_{2}}\mathcal{F}_{1}(\mathcal{F}_{1}-3\mathcal{F}_{2}),\label{app:Chdzdz}\\
\hat{\mathcal{C}}_{\dot{\zeta}\delta\dot{\psi}} & = & -\frac{a^{3}H}{\mathcal{F}_{2}}(\mathcal{F}_{1}-3\mathcal{F}_{2})(\mathcal{F}_{1Z}\dot{\bar{\phi}}+2\mathcal{F}_{1Y}\dot{\bar{\psi}}),\label{app:Chdzdp}
\end{eqnarray}
\begin{eqnarray}
\hat{\mathcal{C}}_{\delta\dot{\psi}\delta\dot{\psi}} & = & -\frac{a^{3}H^{2}}{16\mathcal{F}_{2}\mathcal{F}_{4}}\Big\{9\mathcal{F}_{1XZ}{}^{2}\mathcal{F}_{2}\dot{\bar{\phi}}^{6}+18(\mathcal{F}_{1ZZ}+2\mathcal{F}_{1XY})\mathcal{F}_{1XZ}\mathcal{F}_{2}\dot{\bar{\phi}}^{5}\dot{\bar{\psi}}\nonumber \\
 &  & \quad+9\mathcal{F}_{2}\dot{\bar{\phi}}^{4}\Big[6\mathcal{F}_{1XZ}\mathcal{F}_{1Z}+(\mathcal{F}_{1ZZ}{}^{2}+4\mathcal{F}_{1ZZ}\mathcal{F}_{1XY}+4\mathcal{F}_{1XY}{}^{2}+6\mathcal{F}_{1YZ}\mathcal{F}_{1XZ})\dot{\bar{\psi}}^{2}\Big]\nonumber \\
 &  & \quad+18\mathcal{F}_{2}\dot{\bar{\phi}}^{3}\Big[3\bigl(2\mathcal{F}_{1Y}\mathcal{F}_{1XZ}+(\mathcal{F}_{1ZZ}+2\mathcal{F}_{1XY})\mathcal{F}_{1Z}\bigr)\dot{\bar{\psi}}+(3\mathcal{F}_{1ZZ}\mathcal{F}_{1YZ}+6\mathcal{F}_{1YZ}\mathcal{F}_{1XY}+2\mathcal{F}_{1YY}\mathcal{F}_{1XZ})\dot{\bar{\psi}}^{3}\Big]\nonumber \\
 &  & \quad+\dot{\bar{\phi}}^{2}\Big[-6\mathcal{F}_{1ZZ}\mathcal{F}_{2}\mathcal{F}_{4}+\mathcal{F}_{1Z}^{2}(81\mathcal{F}_{2}+4\mathcal{F}_{4})+54(2\mathcal{F}_{1ZZ}\mathcal{F}_{1Y}+4\mathcal{F}_{1Y}\mathcal{F}_{1XY}+3\mathcal{F}_{1YZ}\mathcal{F}_{1Z})\mathcal{F}_{2}\dot{\bar{\psi}}^{2}\nonumber \\
 &  & \qquad+9\bigl(9\mathcal{F}_{1YZ}{}^{2}+4\mathcal{F}_{1YY}(\mathcal{F}_{1ZZ}+2\mathcal{F}_{1XY})\bigr)\mathcal{F}_{2}\dot{\bar{\psi}}^{4}\Big]\nonumber \\
 &  & \quad+4\Big[-6\mathcal{F}_{1Y}\mathcal{F}_{2}\mathcal{F}_{4}+(81\mathcal{F}_{1Y}\mathcal{F}_{1Z}\mathcal{F}_{2}+4\mathcal{F}_{1Y}\mathcal{F}_{1Z}\mathcal{F}_{4}-6\mathcal{F}_{1YZ}\mathcal{F}_{2}\mathcal{F}_{4})\dot{\bar{\phi}}\dot{\bar{\psi}}\nonumber \\
 &  & \qquad+\Big(-6\mathcal{F}_{1YY}\mathcal{F}_{2}\mathcal{F}_{4}+\mathcal{F}_{1Y}{}^{2}(81\mathcal{F}_{2}+4\mathcal{F}_{4})\Big)\dot{\bar{\psi}}^{2}+27(3\mathcal{F}_{1Y}\mathcal{F}_{1YZ}+\mathcal{F}_{1YY}\mathcal{F}_{1Z})\mathcal{F}_{2}\dot{\bar{\phi}}\dot{\bar{\psi}}^{3}\nonumber \\
 &  & \qquad+54\mathcal{F}_{1YY}\mathcal{F}_{1Y}\mathcal{F}_{2}\dot{\bar{\psi}}^{4}+27\mathcal{F}_{1YY}\mathcal{F}_{1YZ}\mathcal{F}_{2}\dot{\bar{\phi}}\dot{\bar{\psi}}^{5}+9\mathcal{F}_{1YY}{}^{2}\mathcal{F}_{2}\dot{\bar{\psi}}^{6}\Big]\Big\},\label{app:Chdpdp}
\end{eqnarray}
\begin{eqnarray}
\hat{\mathcal{C}}_{\partial\delta\dot{\psi}\partial\delta\dot{\psi}} & = & \frac{a}{4\mathcal{F}_{4}\dot{\bar{\psi}}^{2}}\Big[4f_{9}\mathcal{F}_{4}-9(f_{8}+2f_{9})\mathcal{F}_{1Z}\dot{\bar{\phi}}\dot{\bar{\psi}}-3(f_{8}+2f_{9})\mathcal{F}_{1XZ}\dot{\bar{\phi}}^{3}\dot{\bar{\psi}}-3(f_{8}+2f_{9})(\mathcal{F}_{1ZZ}+2\mathcal{F}_{1XY})\dot{\bar{\phi}}^{2}\dot{\bar{\psi}}^{2}\nonumber \\
 &  & \quad-6(f_{8}+2f_{9})\mathcal{F}_{1YY}\dot{\bar{\psi}}^{4}-9(f_{8}+2f_{9})\dot{\bar{\psi}}^{2}(2\mathcal{F}_{1Y}+\mathcal{F}_{1YZ}\dot{\bar{\phi}}\dot{\bar{\psi}})\Big],
\end{eqnarray}
\begin{equation}
\hat{\mathcal{C}}_{\partial^{2}\delta\dot{\psi}\partial^{2}\delta\dot{\psi}}=-\frac{(f_{8}+2f_{9})^{2}}{4\mathcal{F}_{4}aH^{2}\dot{\bar{\psi}}^{2}},
\end{equation}
\begin{equation}
\hat{\mathcal{C}}_{\zeta\delta\dot{\psi}}=\frac{1}{2\mathcal{F}_{2}}(\mathcal{F}_{1}-3\mathcal{F}_{2})a\Big(4\mathcal{F}_{1\psi}a^{2}H-3a^{2}H^{2}(\mathcal{F}_{1Z}\dot{\bar{\phi}}+2\mathcal{F}_{1Y}\dot{\bar{\psi}})\Big),
\end{equation}
\begin{eqnarray}
\hat{\mathcal{C}}_{\partial^{2}\zeta\delta\dot{\psi}} & = & \frac{a}{\mathcal{F}_{2}\mathcal{F}_{4}\dot{\bar{\psi}}}\Big\{-(f_{3}\mathcal{F}_{1}+f_{4}\mathcal{F}_{1}+2f_{5}\mathcal{F}_{1}+2f_{6}\mathcal{F}_{1}-4f_{11}\mathcal{F}_{2}-\mathcal{F}_{2}\mathcal{F}_{3})\mathcal{F}_{4}+3(f_{10X}+f_{11X})\mathcal{F}_{1XZ}\mathcal{F}_{2}\dot{\bar{\phi}}^{5}\dot{\bar{\psi}}\nonumber \\
 &  & \quad+3\Big[f_{10X}(\mathcal{F}_{1ZZ}+2\mathcal{F}_{1XY})+f_{11X}(\mathcal{F}_{1ZZ}+2\mathcal{F}_{1XY})+(f_{10Z}+f_{11Z})\mathcal{F}_{1XZ}\Big]\mathcal{F}_{2}\dot{\bar{\phi}}^{4}\dot{\bar{\psi}}^{2}\nonumber \\
 &  & \qquad+9(2f_{10Z}\mathcal{F}_{1Y}+2f_{11Z}\mathcal{F}_{1Y}-f_{10}\mathcal{F}_{1YZ}-f_{11}\mathcal{F}_{1YZ}+f_{10Y}\mathcal{F}_{1Z}+f_{11Y}\mathcal{F}_{1Z})\mathcal{F}_{2}\dot{\bar{\phi}}\dot{\bar{\psi}}^{3}\nonumber \\
 &  & \quad-6\Big[f_{10}\mathcal{F}_{1YY}+f_{11}\mathcal{F}_{1YY}-3(f_{10Y}+f_{11Y})\mathcal{F}_{1Y}\Big]\mathcal{F}_{2}\dot{\bar{\psi}}^{4}\nonumber \\
 &  & \quad+3\Big[2f_{10Z}\mathcal{F}_{1YY}+2f_{11Z}\mathcal{F}_{1YY}+3(f_{10Y}+f_{11Y})\mathcal{F}_{1YZ}\Big]\mathcal{F}_{2}\dot{\bar{\phi}}\dot{\bar{\psi}}^{5}\nonumber \\
 &  & \quad+6(f_{10Y}+f_{11Y})\mathcal{F}_{1YY}\mathcal{F}_{2}\dot{\bar{\psi}}^{6}-\mathcal{F}_{2}\dot{\bar{\psi}}\Big[\Big(9f_{10}\mathcal{F}_{1Z}+9f_{11}\mathcal{F}_{1Z}+2(f_{10Z}+f_{11Z})\mathcal{F}_{4}\Big)\dot{\bar{\phi}}\nonumber \\
 &  & \qquad+2\Big(9f_{10}\mathcal{F}_{1Y}+9f_{11}\mathcal{F}_{1Y}+2(f_{10Y}+f_{11Y})\mathcal{F}_{4}\Big)\dot{\bar{\psi}}\Big]\nonumber \\
 &  & \quad+3\mathcal{F}_{2}\dot{\bar{\phi}}^{3}\dot{\bar{\psi}}\Big[-f_{10}\mathcal{F}_{1XZ}-f_{11}\mathcal{F}_{1XZ}+3f_{10X}\mathcal{F}_{1Z}+3f_{11X}\mathcal{F}_{1Z}\nonumber \\
 &  & \qquad+\Big(f_{11Z}\mathcal{F}_{1ZZ}+3f_{10X}\mathcal{F}_{1YZ}+3f_{11X}\mathcal{F}_{1YZ}+2f_{11Z}\mathcal{F}_{1XY}+f_{10Z}(\mathcal{F}_{1ZZ}+2\mathcal{F}_{1XY})\nonumber \\
 &  & \qquad+f_{10Y}\mathcal{F}_{1XZ}+f_{11Y}\mathcal{F}_{1XZ}\Big)\dot{\bar{\psi}}^{2}\Big]+3\mathcal{F}_{2}\dot{\bar{\phi}}^{2}\dot{\bar{\psi}}\bigg(-\Big[f_{10}(\mathcal{F}_{1ZZ}+2\mathcal{F}_{1XY})\nonumber \\
 &  & \qquad+f_{11}(\mathcal{F}_{1ZZ}+2\mathcal{F}_{1XY})-3\Big(2f_{10X}\mathcal{F}_{1Y}+2f_{11X}\mathcal{F}_{1Y}+(f_{10Z}+f_{11Z})\mathcal{F}_{1Z}\Big)\Big]\dot{\bar{\psi}}\nonumber \\
 &  & \qquad+(2f_{10X}\mathcal{F}_{1YY}+2f_{11X}\mathcal{F}_{1YY}+f_{10Y}\mathcal{F}_{1ZZ}+f_{11Y}\mathcal{F}_{1ZZ}+3f_{10Z}\mathcal{F}_{1YZ}\nonumber \\
 &  & \qquad+3f_{11Z}\mathcal{F}_{1YZ}+2f_{10Y}\mathcal{F}_{1XY}+2f_{11Y}\mathcal{F}_{1XY})\dot{\bar{\psi}}^{3}\bigg)\Big\},
\end{eqnarray}
\begin{equation}
\hat{\mathcal{C}}_{\partial^{2}\zeta\partial^{2}\delta\dot{\psi}}=\frac{2(f_{8}+2f_{9})}{\mathcal{F}_{4}aH^{2}\dot{\bar{\psi}}}\Big[f_{10}+f_{11}-(f_{10X}+f_{11X})\dot{\bar{\phi}}^{2}-\dot{\bar{\psi}}\Big((f_{10Z}+f_{11Z})\dot{\bar{\phi}}+(f_{10Y}+f_{11Y})\dot{\bar{\psi}}\Big)\Big].
\end{equation}

The determinant of the Hessian matrix of $\dot{\zeta}$ and $\delta\dot{\psi}$
can be obtained as
\begin{eqnarray}
\det\bm{H} & = & 4\hat{\mathcal{C}}_{\dot{\zeta}\dot{\zeta}}\hat{\mathcal{C}}_{\delta\dot{\psi}\delta\dot{\psi}}-(\hat{\mathcal{C}}_{\dot{\zeta}\delta\dot{\psi}})^{2}\nonumber \\
 & = & \frac{3(\mathcal{F}_{1}-3\mathcal{F}_{2})a^{6}H^{2}}{4\mathcal{F}_{2}\mathcal{F}_{4}}\Big\{3\mathcal{F}_{1}\mathcal{F}_{1XZ}{}^{2}\dot{\bar{\phi}}^{6}+6\mathcal{F}_{1}(\mathcal{F}_{1ZZ}+2\mathcal{F}_{1XY})\mathcal{F}_{1XZ}\dot{\bar{\phi}}^{5}\dot{\bar{\psi}}\nonumber \\
 &  & \quad+3\mathcal{F}_{1}\dot{\bar{\phi}}^{4}\Big[6\mathcal{F}_{1XZ}\mathcal{F}_{1Z}+(\mathcal{F}_{1ZZ}{}^{2}+4\mathcal{F}_{1ZZ}\mathcal{F}_{1XY}+4\mathcal{F}_{1XY}{}^{2}+6\mathcal{F}_{1YZ}\mathcal{F}_{1XZ})\dot{\bar{\psi}}^{2}\Big]\nonumber \\
 &  & \quad+6\mathcal{F}_{1}\dot{\bar{\phi}}^{3}\Big[3\Big(2\mathcal{F}_{1Y}\mathcal{F}_{1XZ}+(\mathcal{F}_{1ZZ}+2\mathcal{F}_{1XY})\mathcal{F}_{1Z}\Big)\dot{\bar{\psi}}+(3\mathcal{F}_{1ZZ}\mathcal{F}_{1YZ}+6\mathcal{F}_{1YZ}\mathcal{F}_{1XY}+2\mathcal{F}_{1YY}\mathcal{F}_{1XZ})\dot{\bar{\psi}}^{3}\Big]\nonumber \\
 &  & \qquad+\dot{\bar{\phi}}^{2}\Big[27\mathcal{F}_{1}\mathcal{F}_{1Z}^{2}-2\mathcal{F}_{1}\mathcal{F}_{1ZZ}\mathcal{F}_{4}+4\mathcal{F}_{1Z}^{2}\mathcal{F}_{4}+18\mathcal{F}_{1}(2\mathcal{F}_{1ZZ}\mathcal{F}_{1Y}+4\mathcal{F}_{1Y}\mathcal{F}_{1XY}+3\mathcal{F}_{1YZ}\mathcal{F}_{1Z})\dot{\bar{\psi}}^{2}\nonumber \\
 &  & \hspace*{4em}+3\mathcal{F}_{1}\Big(9\mathcal{F}_{1YZ}{}^{2}+4\mathcal{F}_{1YY}(\mathcal{F}_{1ZZ}+2\mathcal{F}_{1XY})\Big)\dot{\bar{\psi}}^{4}\Big]\nonumber \\
 &  & \quad+4\Big[-2\mathcal{F}_{1}\mathcal{F}_{1Y}\mathcal{F}_{4}+(27\mathcal{F}_{1}\mathcal{F}_{1Y}\mathcal{F}_{1Z}-2\mathcal{F}_{1}\mathcal{F}_{1YZ}\mathcal{F}_{4}+4\mathcal{F}_{1Y}\mathcal{F}_{1Z}\mathcal{F}_{4})\dot{\bar{\phi}}\dot{\bar{\psi}}\nonumber \\
 &  & \qquad+(27\mathcal{F}_{1}\mathcal{F}_{1Y}^{2}-2\mathcal{F}_{1}\mathcal{F}_{1YY}\mathcal{F}_{4}+4\mathcal{F}_{1Y}^{2}\mathcal{F}_{4})\dot{\bar{\psi}}^{2}+9\mathcal{F}_{1}(3\mathcal{F}_{1Y}\mathcal{F}_{1YZ}+\mathcal{F}_{1YY}\mathcal{F}_{1Z})\dot{\bar{\phi}}\dot{\bar{\psi}}^{3}\nonumber \\
 &  & \qquad+18\mathcal{F}_{1}\mathcal{F}_{1YY}\mathcal{F}_{1Y}\dot{\bar{\psi}}^{4}+9\mathcal{F}_{1}\mathcal{F}_{1YY}\mathcal{F}_{1YZ}\dot{\bar{\phi}}\dot{\bar{\psi}}^{5}+3\mathcal{F}_{1}\mathcal{F}_{1YY}^{2}\dot{\bar{\psi}}^{6}\Big]\Big\}.
\end{eqnarray}
One should note that by taking into account the propagating speed
and no-ghost conditions for the GWs, the combination $\mathcal{F}_{1}-3\mathcal{F}_{2}=-4\mathcal{G}<0$
and thus never vanishes. As a result, the Hessian matrix generally
does not degenerate as long as $\mathcal{F}_{1}$ depends on $Y$
or $Z$. 

\newpage{}

\providecommand{\href}[2]{#2}\begingroup\raggedright\endgroup

\end{document}